\documentclass[american,british,english,notitlepage, twocolumn]{revtex4-1}
\usepackage[T1]{fontenc}
\usepackage[latin9]{inputenc}
\usepackage{amsmath}
\usepackage{amssymb}
\usepackage{mathdots}
\usepackage{graphicx}
\usepackage{esint}

\makeatletter

\def\onehalf{\slantfrac{1}{2}}

\usepackage{hyperref}

\makeatother

\usepackage{babel}
\begin{document}

\title{An Exact Solution of the Fokker-Planck Equation for Isotropic Scattering }

\author{M.A. Malkov}

\affiliation{$^{1}$University of California San Diego, La Jolla, CA 92093}
\begin{abstract}
The Fokker-Planck (FP) equation $\partial_{t}f+\mu\partial_{x}f=\partial_{\mu}\left(1-\mu^{2}\right)\partial_{\mu}f$
is solved analytically. Foremost among its applications, this equation
describes the propagation of energetic particles through a scattering
medium (in $x$- direction, with $\mu$ being the $x$- projection
of particle velocity). The solution is found in terms of an infinite
series of mixed moments of particle distribution, $\left\langle \mu^{j}x^{k}\right\rangle $.
The second moment $\left\langle x^{2}\right\rangle $ ($j=0,\;k=2$)
was obtained by G.I. Taylor (1920) in his classical study of\emph{
random walk}: $\left\langle x^{2}\right\rangle =\left\langle x^{2}\right\rangle _{0}+t/3+\left[\exp\left(-2t\right)-1\right]/6$
(where $t$ is given in units of an average time between collisions).
It characterizes a spatial dispersion of a particle cloud released
at $t=0$, with $\sqrt{\left\langle x^{2}\right\rangle _{0}}$ being
its initial width. This formula distills a transition from ballistic
(rectilinear) propagation phase, $\left\langle x^{2}\right\rangle -\left\langle x^{2}\right\rangle _{0}\approx t^{2}/3$
to a time-asymptotic, diffusive phase, $\left\langle x^{2}\right\rangle -\left\langle x^{2}\right\rangle _{0}\approx t/3$.
The present paper provides all the higher moments by a recurrence
formula. The full set of moments is equivalent to the full solution
of FP equation, expressed in form of an infinite series in moments
$\left\langle \mu^{j}x^{k}\right\rangle $. An explicit, easy-to-use
approximation for a point source spreading of a pitch-angle averaged
distribution $f_{0}\left(x,t\right)$ (starting from $f_{0}\left(x,0\right)=\delta\left(x\right),$
i.e., Green's function), is also presented and verified by a numerical
integration of FP equation.
\end{abstract}
\maketitle

\section{Introduction}

Propagation of energetic particles, which we will call cosmic rays
(or CR for short), particularly through magnetized turbulent media,
has been actively researched in the astrophysical community for more
than half a century. The time asymptotic solution of this problem
is known to be diffusive. After several collisions particles \textquotedblleft forget\textquotedblright{}
their initial velocities and enter a random walk process. However,
in astrophysical objects with infrequent particle collisions, there
is not enough time or space for even a few collisions. In such systems,
an early-time propagation is not random as particles \textquotedblleft remember\textquotedblright{}
their starting velocities and positions. 

At times much shorter than the collision time, $t\ll t_{c}$, most
particles propagate with their initial velocities or their projections
on the magnetic field direction, if present. This regime is called
the ballistic, or rectilinear propagation. The question then is what
happens next, namely at $t\sim t_{c}$ but before the onset of diffusion
at $t\gg t_{c}$? What exactly is the value of $t/t_{c}\gtrsim1$,
when the simple diffusive description becomes applicable? In other
words, what is the extent of an intermediate phase when neither ballistic
nor diffusive model applies? These are the questions we address below
using a Fokker-Planck (FP) transport equation and its exact analytic
solution. 

A review of early results on CR propagation is contained, e.g., in
Refs.\cite{Jok1971RvGSP,Voelk73}, while Ref.\cite{M_PoP2015} covers
some recent results relevant to the present paper. The FP  model is
extensively applied by the CR and heliophysics communities to a wide
range of transport processes driven by small-step stochastic variations
of particle velocity. This formulation is relevant, e.g., to the solar
wind, through which solar energetic particles and CRs propagate to
the Earth, and interstellar medium, through which CRs propagate from
more distant sources. Under ``collisions'' in such media, one usually
understands particle scattering off magnetic disturbances. Another
example is the propagation of ultra-high-energy CRs from extragalactic
sources. The transition from ballistic to diffusive transport regime,
while being challenging for the theory, is key to understanding the
nature of such sources. Since the particle mean free path usually
grows with energy, part of their spectrum almost inescapably falls
into a transient category where neither ballistic nor diffusive approximation
applies. For the lack of better terms, we will call these particles
transdiffusive.

Transdiffusive particles are likely to carry most of the information
about their source. Indeed, the low-energy, diffusively propagating
particles either do not reach us in time or merge into a featureless
isotropic background. The highest energy particles, on the contrary,
propagate ballistically. They may point back to their sources and
are therefore invaluable. Unfortunately, they are exceptionally rare.
On average, just one CR particle with energy $>10^{20}$eV is expected
to arrive per century per square kilometer. Only a handful of such
events has been registered over decades of observation. Their number
grows towards lower energies, but their arrival direction becomes
random too, thus making the source indiscernible on the sky. It follows
that it is the transdiffusive propagation regime that may steer particles
between Scylla of poor statistics and Charybdis of orbit scrambling
and unveil the source. Understanding the transdiffusive propagation
is then the key to an emerging field of CR astronomy.

\section{Formulation of the problem and its significance \label{sec:Formulation-of-the}}

As there is no lack of motivation for studying the transdiffusive
propagation regime, it is worthwhile to discuss the Fokker-Planck
(FP) equation as a simple yet adequate mathematical model for this
purpose. The FP equation is general in that it applies to both magnetized
and unmagnetized media, as long as particles (or other entities, such
as wave quanta) scatter randomly at small angles. The magnetic field,
however, conveniently justifies the one-dimensional reduction of more
realistic three-dimensional problems, as particles are usually bound
to the field lines. After averaging out their gyromotion (typically
unimportant) the particle phase space becomes two-dimensional. The
resulting equation for their distribution function $f$ describes
a one-dimensional spatial transport constrained by angular scattering:

\begin{equation}
\frac{\partial f}{\partial t}+v\mu\frac{\partial f}{\partial x}=\frac{\partial}{\partial\mu}\left(1-\mu^{2}\right)D\frac{\partial f}{\partial\mu}.\label{eq:PAscatIntro}
\end{equation}
Here $x$ is the only spatial variable along which the particle concentration
varies (local field direction or another symmetry axis), $\mu$ is
the cosine of particle pitch angle to the $x-$ axis, $v$ is the
magnitude of particle velocity, conserved in interactions with quasi-static
magnetic turbulence. $D$ is the scattering rate (collision frequency). 

A major propagation scenario that Eq.(\ref{eq:PAscatIntro}) handles
well comes about through an instant release of a small particle cloud
into a scattering medium. For instance, galactic supernova remnants
(SNR), widely believed to generate CRs with energies up to $\sim10^{15}$eV,
must accelerate CRs in their shock waves and subsequently release
them into the turbulent interstellar medium \cite{GrenierRev2015}.
Mathematically, the question then is how exactly the pitch-angle averaged
particle distribution propagates along a magnetic flux tube that intersects
the SNR shell. It is highly desirable to achieve the simplicity of
diffusive description (e.g., \cite{Jokipii66} and below) which is
a well-known derivative of Eq.(\ref{eq:PAscatIntro}). As emphasized
earlier, the diffusive treatment is inadequate during the ballistic
and transdiffusive propagation, while the latter is often the key
for probing into the source. During these phases energetic CR protons
(the main species) may reach a nearby molecular cloud, making themselves
visible by interacting with its dense gas \cite{AharDV94}. The CR
protons of lower energies would instead be diffusively confined to
the SNR shell and remain elusive for observers. Another example is
the propagation of solar energetic particles to 1 AU. Also, in this
case, the particle mean free path (m.f.p.) may be comparable to or
exceed 1 AU, so the diffusion approximation fails again \footnote{Note, however, that if the pressure of released particles is of the
order or larger than the magnetic pressure of the ambient medium,
the particles are self-confined by scattering off self-generated MHD
waves. This kind of problems should be treated differently, \cite{MDS10PPCF,MetalEsc13,NavaGabici2016,DAngeloBlasi2016}
from what is described further in this paper, as the scattering frequency
$D$ would strongly depend on $\nabla f$, whereas here we consider
the case $D=D\left(v\right)$.}. 

\subsection{Simple limiting cases\label{subsec:Simple-limiting-cases}}

Before proceeding with the solution of the FP equation (\ref{eq:PAscatIntro})
in Sec.\ref{sec:Fokker-Planck-Solution}, it is useful to characterize
its limiting cases of ballistic and diffusive propagation. We deduce
them directly from Eq.(\ref{eq:PAscatIntro}), by eliminating angular
dynamics. 

\subsubsection{Ballistic propagation regime\label{subsec:Ballistic-propagation-regime}}

In the ballistic regime which strictly applies to times shorter than
the collision time, $t\ll t_{c}\sim1/D$, one can neglect the r.h.s.
of the equation, and obtain the solution by integrating along the
particle trajectories (Liouville's theorem), $x-\mu vt=const$ with
a conserved pitch angle, $\mu=const$. The solution is simply $f\left(x,\mu,t\right)=f\left(x-v\mu t,\mu,0\right)$. 

It is sufficient to consider here a point source with initially isotropic
distribution: $f\left(x,\mu,0\right)=\onehalf\delta\left(x\right)\Theta\left(1-\mu^{2}\right),$
where $\delta$ and $\Theta$ denote the Dirac's delta and Heaviside
unit step functions, respectively. From the above solution for $f\left(x,\mu,t\right)$,
one obtains the ballistic expansion in form of the second moment,
$\left\langle x^{2}\right\rangle =v^{2}t^{2}/3$ by integrating $x^{2}f=\onehalf x^{2}\delta\left(x-v\mu t\right)\Theta\left(1-\mu^{2}\right)$
over $x$ and $\mu$. The result describes a free escape with the
mean square velocity $v/\sqrt{3}$, while the maximum particle velocity
(along $x$) is $v$. The form of the pitch angle averaged particle
distribution, $f_{0}\left(x,t\right)=\left(2vt\right)^{-1}\Theta\left(1-x^{2}/v^{2}t^{2}\right)$,
is characterized by an expanding 'box' of decreasing height. Note
that this result is inconsistent with the solution of the so called
``telegraph'' equation that has been put forward for CR propagation
over the last 50 years and is discussed below at some length. By contrast,
an exact solution of Eq.(\ref{eq:PAscatIntro}) obtained below converges
to the above-described box distribution at $t\ll1$. 

\subsubsection{Diffusive (hyperdiffusive) propagation regime\label{subsec:Diffusive-(hyperdiffusive)-propa}}

The second simple propagation regime is diffusive which dominates
at $t\gg t_{c}\sim1/D$, and is treated in a way opposite to the above-described
ballistic regime, \cite{Jokipii66}. The r.h.s. of Eq.(\ref{eq:PAscatIntro})
is now the leading term, thus implying that the particle distribution
is close to isotropy, $\partial f/\partial\mu\to0$. Working to higher
orders in anisotropic corrections $\sim1/D$, and averaging the equation
over $\mu$, one obtains the following equation for $f_{0}\left(x,t\right)$
\cite{MS2015}

\begin{equation}
\frac{\partial f_{0}}{\partial t}-\kappa_{2}\frac{\partial^{2}f_{0}}{\partial x^{2}}=-\kappa_{4}\frac{\partial^{4}f_{0}}{\partial x^{4}}+\kappa_{6}\frac{\partial^{6}f_{0}}{\partial x^{6}}-\dots,\label{eq:AppendDiffHyperdiff}
\end{equation}
with $\kappa_{2n}\sim1/D^{n}$. This particular form of expansion
is relevant under the scattering symmetry: $D\left(-\mu\right)=D\left(\mu\right)$.
Otherwise, also the odd $x-$ derivatives appear on its r.h.s. The
latter situation is not considered here for simplicity. The last equation
results from an asymptotic (Chapman-Enskog) expansion of the problem
in $1/D$. It is valid only for $t\gg t_{c}\sim1/D$, and all the
short-time-scale, ballistic propagation effects are intentionally
eliminated (cf. elimination of secular terms in perturbative treatments).
A failure to do so results in a second order time derivative in Eq.(\ref{eq:AppendDiffHyperdiff})
(``telegraph'' term) which is illegitimate unless $t\gg t_{c}$
(see \cite{MS2015,M_PoP2015} and below). Meanwhile, the r.h.s. of
the above equation provides a small hyperdiffusive correction which
may be omitted, as the higher spatial derivatives quickly decay because
of the smoothing effect from the diffusive term on its l.h.s. 

The most serious problem with the diffusive approximation is an unrealistically
high propagation speed of particles which reach a given point faster
than the maximum speed would allow (acausal solution, e.g. \cite{SchwadronTelegraph94,Berezinsky2007,AloisBerezSuperLum09}).
Mathematically, the approximation violates an upper bound $\left|x\right|\le vt$
that immediately follows from Eq.(\ref{eq:PAscatIntro}) for a point
source solution. There have been many attempts to overcome this problem,
but no adequate \emph{ab initio} description of particle spreading
that would cover ballistic and diffusive phases was elaborated. Much
popularity received the telegraph equation, mentioned above. Ian Axford
\cite{AxfordTelegr65} derived it semi-empirically and supported by
studies of \emph{discontinuous }random walk \cite{GOLDSTEIN01011951}
back in 1965. Many other derivations of telegraph equation from Eq.(\ref{eq:PAscatIntro})
and attempts to justify its application to various CR transport problems
have appeared ever since, e.g., \cite{Earl1974,SchwadronTelegraph94,LitvinEffenSchlick15,Litvin2016PhPl}.
However, the solution of the telegraph equation is arguably inconsistent
with the FP equation. It is sufficient to mention a presence of singular
components incompatible with the parent equation, eq.{[}\ref{eq:PAscatIntro}{]}.
These are introduced to make up for nonconservation of the total number
of particles which we also discuss below.

Both diffusive and \textquotedblleft telegraph\textquotedblright{}
approaches are aimed to extract the spatial particle distribution
from Eq.(\ref{eq:PAscatIntro}). The elimination of pitch angle, however,
has always been a challenge. While the time-asymptotic character of
the diffusive approach is well understood, the idea behind the telegraph
equation was to span also the transdiffusive evolution, preferably
down to the ballistic phase. As we mentioned, the derivation of telegraph
equation is inconsistent with a regular Chapman-Enskog asymptotic
expansion (systematic elimination of pitch angle) and the ballistic
propagation. The higher order Chapman-Enskog expansion, in turn, is
valid only for time-asymptotic regimes, $t\gg t_{c}$, and should
be considered as a correction to the diffusive treatment. As in many
other asymptotic expansions, when applied outside of their validity
range, higher order terms often make the approximation less accurate
which was recently demonstrated in Ref.\foreignlanguage{american}{\cite{Litvin2016PhPl}},
using the numerical integration of Eq.(\ref{eq:PAscatIntro}). By
contrast to the hyperdiffusive Chapman-Enskog expansion, the telegraph
equation, as mentioned above, was intended to cover also the crossover
phase, $t\sim t_{c}$. Not surprisingly, its solution failed to provide
an adequate fit to a full numerical solution, thus confirming its
failure at $t\lesssim t_{c}$.

Being most easily obtained from the hyperdiffusive corrections in
Eq.(\ref{eq:AppendDiffHyperdiff}), the telegraph equation inherits
its validity range, $t\gg t_{c}$. In this appearance, it constitutes
just another form of small correction to diffusion. Indeed, considering
the two terms on the l.h.s. of Eq.(\ref{eq:AppendDiffHyperdiff})
as leading (which is required by its derivation!) and converting then
the fourth spatial derivative (the leading term on the r.h.s) into
a second time-derivative, we write: $f_{0xxxx}\simeq f_{0tt}/\kappa_{2}^{2}$.
By dropping higher $x$- derivatives one recovers the telegraph equation 

\begin{equation}
\frac{\partial^{2}f_{0}}{\partial t^{2}}-V^{2}\frac{\partial^{2}f_{0}}{\partial x^{2}}+\tau^{-1}\frac{\partial f_{0}}{\partial t}=0.\label{eq:telegraph}
\end{equation}
At first glimpse, it indeed captures a ballistic (wave-like) propagation
of particle bunches at a speed $V=\sqrt{\kappa_{2}^{3}/\kappa_{4}}$.
However, the number density in the bunches decays with time at a rate
$\tau^{-1}=\kappa_{2}^{2}/\kappa_{4}$ which is nonphysical. Unlike
eqs.(\ref{eq:PAscatIntro}) and (\ref{eq:AppendDiffHyperdiff}), this
equation does not conserve the number of particles, $N=\int f_{0}dx$,
automatically. To conserve the total number of particles in a fundamental
solution (Green's function) for Eq.(\ref{eq:telegraph}), two different
solutions need to be added together. As the equation is linear, such
addition is indeed legitimate. The first component of the solution
is smooth within the characteristics of eq(\ref{eq:telegraph}), $\left|x\right|<Vt$
and zero otherwise, thus it develops discontinuities at $x=\pm Vt$.
The number of particles contained in this solution component grows
in time from zero as they start spreading ballistically along the
characteristics. The second solution, whose sole purpose is to compensate
for the nonphysical multiplication of particles in the first component,
needs to be taken in even more singular form, $f_{a}=\onehalf N\left(t\right)\left[\delta\left(x-Vt\right)+\delta\left(x+Vt\right)\right]$.
The number of particles in this component decays as $N=N\left(0\right)\exp\left(-t/2\tau\right)$.
However, this auxiliary component is inconsistent with the original
equation (\ref{eq:PAscatIntro}). Indeed, the particle distribution
of the form$\propto\delta\left(x\pm Vt\right)$ implies that all particles
have the same pitch angle. But the operator of angular scattering
on the r.h.s. of Eq.(\ref{eq:PAscatIntro}) would momentarily smear
out any sharp pitch-angle distribution. Therefore, the $\delta\left(x\pm Vt\right)$
components cannot persist in the parent FP equation and have been
illegitimately acquired during its reduction \footnote{Evidently, the above arguments do not apply to a telegraph equation
derived from a Boltzmann equation with a simplified scattering operator
of the form $St\left(f\right)=\left(f_{0}-f\right)/\tau$ (so called
BGK, or $\tau-$ approximation). In this context, the telegraph equation
has been studied in, e.g., \cite{Gombosi93,Zank2000,Webb2000,WebbTelegr06,Fedorov2016}.}. 

We conclude that the telegraph equation can be accepted only at $t\gg t_{c}$
when the nonphysical $\delta$- peaks die out. The telegraph equation
can then be deduced from the hyperdiffusive expansion in Eq.(\ref{eq:AppendDiffHyperdiff}),
obtained, in turn, by using the canonical Chapman-Enskog approach.
At earlier times, the telegraph solution does not match the actual
solution of Eq.(\ref{eq:PAscatIntro}), e.g., \cite{Litvin2016PhPl}.
Interestingly enough, only at $t\gtrsim10t_{c}$ merges the telegraph
solution with diffusive, hyperdiffusive and direct numerical solution
of the parent FP equation. Even then, nonphysical peaks stemming from
the singular part of the solution remain well pronounced (Fig.5 in
the above paper). 

Notwithstanding the irrelevance of the telegraph solution to early
phases of particle transport, singular $\delta$- components do arise
in a different context of the telegraph equation (apart from transmission
lines, of course). It is associated with a \emph{discontinuous} random
walk. In this process, studied in Ref. \cite{GOLDSTEIN01011951} and
earlier by G.I. Taylor \cite{Taylor01011922}, particles are allowed
to move only at fixed velocities, positive or negative, say $\pm V$,
which naturally results in a $\delta\left(x\pm Vt\right)$ particle
distribution. Statistically, this is similar to a coin tossing with
only two possible outcomes. Under a continuous pitch-angle dependence
relevant to the CR propagation, the discontinuous random walk restriction
corresponds to the particle distribution, concentrated at $\mu=\pm1$,
thus producing the $\delta\left(x\pm Vt\right)$ singular components
of the telegraph equation. As we noted, however, the underlying $\delta\left(\mu\pm1\right)$
angular distribution is inconsistent with the solutions of FP Eq.(\ref{eq:PAscatIntro}).
That is why the telegraph equation cannot be consistently derived
from the FP equation except for $t\gg t_{c}$. Being alternatively
derived from the discontinuous random walk process, it is unsuitable
for describing the transport of energetic particles since their spectrum
in velocity projection on the travel direction (pitch angle) is a
fundamentally continuous variable.

It follows that apart from the well established diffusive description
of Eq.(\ref{eq:PAscatIntro}), though valid only for $t\gg t_{c}\sim1/D$,
there are no viable analytical tools to address the earlier phases
of particle propagation. Therefore, an exact solution of the FP equation
we tackle below is more than motivated. 

\section{Fokker-Planck Equation and its Solution\label{sec:Fokker-Planck-Solution}}

Turning to the solution of Fokker-Planck Eq.(\ref{eq:PAscatIntro}),
we first discuss its scope. The energy dependence of the particle
scattering frequency enters it only as a \emph{parameter}, i.e., $D\left(E\right)$,
and does not prevent us from solving the equation. By contrast, a
possible pitch-angle dependence of $D\left(\mu\right)$ (nonisotropic
scattering) does. However, in media with magnetic irregularities,
$D\left(\mu\right)$ derives from a power index of the scattering
turbulence, $q$, if the interaction between particles and turbulence
is resonant, e.g. \cite{VVSQL62,Kennel66,Jokipii66,Skill75a,BlandEich87}.
In particular, for a power spectrum $P\propto k^{-q}$, where $k$
is the wave number, one obtains $D\left(\mu\right)\propto\left|\mu\right|^{q-1}$
and the scattering is isotropic for an important case $q=1$, as discussed
below. Even more complex, anisotropic turbulence spectra, such as
those derived by Goldreich and Shridhar \cite{goldr97}, result in
a flat $D\left(\mu\right)$ \cite{ChandranGS00PhRvL}, except for
relatively narrow regions near $\mu=0,$$\pm1$. These areas require
special considerations in any event, as they are strongly affected
by particle mirroring and resonance broadening \cite{Acht81a} ($\mu\approx0$),
as well as by the field aligned propagation \cite{Milagro10} ($\left|\mu\right|\approx1)$.
We may ignore them as they occupy only a small fraction of particle
phase space. Besides, fluctuation spectra with the index $q\approx1$
have been obtained in Monte-Carlo studies of shock-accelerated particles
\cite{Bykov2014ApJ}. The choice of $\mu$-independent scattering
coefficient $D$ has been advocated in \cite{Shalchi2009} even for
strong magnetic fluctuations, $\delta B\gtrsim B_{0}$, where $B_{0}$
is the unperturbed field. Motivated by the above, we consider the
case $D\left(\mu,E\right)=D\left(E\right)$ as it embraces many physically
interesting situations and, at the same time, allows for an exact
analytic solution of Eq.(\ref{eq:PAscatIntro}).

We now rewrite Eq.(\ref{eq:PAscatIntro}) using dimensionless time
and length units according to the following transformations

\begin{equation}
D\left(E\right)t\to t,\;\;\;\frac{D}{v}x\to x\label{eq:RescaledTimeANDx}
\end{equation}
Instead of Eq.(\ref{eq:PAscatIntro}) we thus have

\begin{equation}
\frac{\partial f}{\partial t}+\mu\frac{\partial f}{\partial x}=\frac{\partial}{\partial\mu}\left(1-\mu^{2}\right)\frac{\partial f}{\partial\mu}\label{eq:FPundim}
\end{equation}
This equation contains no explicit parameters thus precluding any
uniformly valid asymptotic expansion unless a small parameter enters
the problem implicitly through the initial condition $f\left(x,\mu,0\right)$.
In particular, if one is interested in an isotropic approximation
that can be treated using Eq.(\ref{eq:AppendDiffHyperdiff}), ($1/D$-
type expansion), not only should the initial distribution be close
to isotropy, $\partial f/\partial\mu\ll f$, but it should also be
spatially broad, $f^{-1}\left|\partial f/\partial x\right|\ll1$.
The latter condition prevents a high anisotropy from arising in the
course of time via the second term on the l.h.s. of Eq.(\ref{eq:FPundim}).
Hence, the fundamental problem of a point source spreading (Green's
function, or fundamental solution) can not be treated using a conventional
$1/D$ expansion, until $f$ becomes quasi-isotropic, that is broadened
to $x\gtrsim1$. This is another reason why the telegraph equation
falls short in describing CR propagation from a point source. 

Based on the above considerations, we tackle an exact solution of
Eq.(\ref{eq:FPundim}). The only restriction that we impose on the
spatial distribution at $t=0$, which holds up during its subsequent
evolution, is a sufficiently rapid decay of $f\left(x\right)$ at
$\left|x\right|\to\infty$. Namely, we require that $x^{n}f\left(x\right)\to0$
for $\left|x\right|\to\infty$ and $n\geq0$. This standard restriction
guarantees the existence of all moments. Another standart restriction
is the regularity of $f$ at $\left|\mu\right|=1$: $\left(1-\mu^{2}\right)f\to0$
for $\left|\mu\right|\to1$. 

Turning to the solution of Eq.(\ref{eq:FPundim}) we introduce moments
of $f\left(\mu,x\right)$ in the form of the following matrix

\begin{equation}
M_{ij}\left(t\right)=\left\langle \mu^{i}x^{j}\right\rangle =\int_{-\infty}^{\infty}dx\int_{-1}^{1}\mu^{i}x^{j}fd\mu/2\label{eq:MomentsDef}
\end{equation}
for any integer $i,j\ge0$. We will discuss conditions for the equivalence
of the moments $M$ and the distribution $f$ when the solution for
the matrix $M$ is obtained. 

The lowest moment $M_{00}$ is automatically conserved by Eq.(\ref{eq:FPundim})
(as being proportional to the number of particles) and we normalize
it to unity 

\[
M_{00}=\int_{-\infty}^{\infty}dx\int_{-1}^{1}fd\mu/2=1.
\]
Multiplying Eq.(\ref{eq:FPundim}) by $\mu^{i}x^{j}$ and integrating
by parts (where appropriate), we obtain the following matrix equation
for the moments $M_{ij}$ with $i+j>0$:

\begin{equation}
\frac{d}{dt}M_{ij}+i\left(i+1\right)M_{ij}-jM_{i+1,j-1}=i\left(i-1\right)M_{i-2,j}\label{eq:Mijdot}
\end{equation}
This equation couples triads of matrix elements. The two elements
are on the same antidiagonal (l.h.s), and the third one is on the
next but one antidiagonal above it (r.h.s). One may get an impression
that this infinite system of moment equations will require truncation
to become useful. Actually, not only is this unnecessary but should
be avoided at all cost, as we argue below. The set of moments $M_{ij}\left(t\right)$
in Eq.(\ref{eq:Mijdot}) can, in fact, be recursively obtained to
any order $n=i+j$ with no truncation. Indeed, as $M_{00}=1,$ and
we can set $M_{ik}=M_{ki}=0$ for any $i<0$, $k\ge0,$ all the elements
of the infinite matrix

\[
M=\left(\begin{array}{ccccc}
1 & \left\langle x\right\rangle  & \left\langle x^{2}\right\rangle  & \left\langle x^{3}\right\rangle \\
\left\langle \mu\right\rangle  & \left\langle \mu x\right\rangle  & \left\langle \mu x^{2}\right\rangle  & \nearrow\\
\left\langle \mu^{2}\right\rangle  & \left\langle \mu^{2}x\right\rangle  & \nearrow & \iddots\\
\left\langle \mu^{3}\right\rangle  & \nearrow & \iddots\\
\nearrow & \iddots
\end{array}\right)
\]
on each antidiagonal can be found, working from left to right as shown
by the arrows. Alternatively, the matrix of moments can be viewed
as a triangle (akin to Pascal's or Bernoulli's triangles, for example)
with its rows being the matrix antidiagonals, starting from the unity
at the top of the triangle. The two moments on the next antidiagonal
(triangle row), are easily found from Eq.(\ref{eq:Mijdot}) to be
$M_{10}\left(t\right)=\left\langle \mu\right\rangle =\left\langle \mu\right\rangle _{0}\exp\left(-2t\right)$
and $M_{01}=\left\langle x\right\rangle =\left\langle x\right\rangle _{0}+\onehalf$$\left\langle \mu\right\rangle _{0}\left[1-\exp\left(-2t\right)\right].$
Higher moments can be obtained inductively. So, in general, from Eq.(\ref{eq:Mijdot})
we find

\begin{equation}
\begin{split}M_{ij}\left(t\right)=M_{ij}\left(0\right)e^{-i\left(i+1\right)t}+\int_{0}^{t}e^{i\left(i+1\right)\left(t^{\prime}-t\right)} & \times\\
\left[jM_{i+1,j-1}\left(t^{\prime}\right)+i\left(i-1\right)M_{i-2,j}\left(t^{\prime}\right)\right]dt^{\prime}
\end{split}
\label{eq:MomSol}
\end{equation}
As may be seen from the above expressions for $M_{01}$ and $M_{10},$
all the higher moments in Eq.(\ref{eq:MomSol}) are series in $t^{k}e^{-nt}$,
where $k$ and $n$ are integral numbers. In particular, the next
set of moments is on the third anti-diagonal, $M_{20},\;M_{11},\;M_{02}$:

\begin{equation}
\begin{split}M_{20}=\frac{1}{3},\;\;\;M_{11}=\frac{1}{6}\left(1-e^{-2t}\right)\\
M_{02}=M_{02}\left(0\right)+\frac{t}{3}-\frac{1}{6}\left(1-e^{-2t}\right)
\end{split}
\label{eq:M2s}
\end{equation}
Interesting in a point source (fundamental) solution, we assume the
initial distribution $f\left(x,\mu,0\right)$ to be symmetric in $x$
and isotropic which eliminates the odd moments. Furthermore, the initial
spatial width must then also be set to zero, $M_{02}\left(0\right)=\left\langle x^{2}\right\rangle _{0}=0$.
Note that $M_{02}\left(t\right)$ in Eq.(\ref{eq:M2s}) coincides
with the respective random walk result obtained by G.I. Taylor \cite{Taylor01011922},
which we discussed earlier. However useful for understanding the transition
between ballistic and diffusive phases of particle propagation, $M_{02}$
\emph{alone} does not, of course, resolve the FP equation. From the
mathematical point of view, only a \emph{full} set of moments in Eq.(\ref{eq:MomSol})
provides a complete solution $f\left(x,\mu,t\right)$ of Eq.(\ref{eq:FPundim})
given the initial value, $f\left(x,\mu,0\right)$ that determines
the matrix $M_{ij}\left(0\right)$ in Eq.(\ref{eq:MomSol}). Moreover,
to adequately reproduce the ballistic and transdiffusive phases the
series of moments cannot be truncated.

In general, the equivalence between an arbitrary distribution $f\left(x,\mu,t\right)$
and its \emph{full} set of moments $M_{ij}\left(t\right)$ is not
guaranteed automatically, but can be established for Eq.(\ref{eq:FPundim})
with its solution in the form of Eq.(\ref{eq:MomSol}) using Hamburger's
theorem, e.g., \cite{reed1975methods}. The theorem assumes upper
bounds on the moments in the form $\left|M_{ij}\right|<An!b^{n}$
with the constants $A$ and $b$ being independent of $n$. According
to Eq.(\ref{eq:MomSol}), for any fixed $t\gg1$ the moments $M_{ij}\left(t\right)$
grow with $n=i+j$ not faster than $t^{n/2}$. Although for small
$t\ll1$ higher powers of $t$ are present, they also have an upper
bound $\sim t^{n}$. Therefore, the condition for Hamburger theorem
is satisfied. 

Being interested in the fundamental solution of the FP equation, we
will focus on the moments that correspond to the isotropic part of
particle distribution 

\begin{equation}
f_{0}\left(x,t\right)=\int_{-1}^{1}f\left(\mu,x,t\right)d\mu/2,\label{eq:f0Def}
\end{equation}
as only this part contributes to the particle number density. Note
that under the fundamental solution we understand here the solution
for $f_{0}\left(x,t\right)$ that it isotropic at $t=0$ and $f_{0}\left(x,0\right)=\delta\left(x\right)$.
It constitutes the pitch-angle averaged distribution and has been
the target of most reduction schemes applied to the FP Eq.(\ref{eq:FPundim}).
The matrix elements that represent $f_{0}$ are, therefore, $M_{0,j}.$
Note, that $M_{ij}$ with $i>0$ are nevertheless not small and remain
essential for calculating the full set of the moments $M_{0,j}$.
To link them to $f_{0}$, we use a standard moment-generating function 

\begin{equation}
f_{\lambda}\left(t\right)=\int_{-\infty}^{\infty}f_{0}\left(x,t\right)e^{\lambda x}dx=\sum_{n=0}^{\infty}\frac{\lambda^{2n}}{\left(2n\right)!}M_{0,2n}\left(t\right)\label{eq:MomGenFunc}
\end{equation}
where we omitted the odd moments irrelevant to the fundamental (symmetric
in $x$) solution. The above expansion will be converted into a Fourier
transform of $f_{0}$$\left(x,t\right)$ by setting $\lambda=-ik$.
To find its inverse, that is the function $f_{0}\left(x,t\right)$,
we will use the inverse Fourier transform

\begin{equation}
f_{0}\left(x,t\right)=\frac{1}{2\pi}\int_{-\infty}^{\infty}dke^{ikx}\sum_{n=0}^{\infty}\frac{\left(-1\right)^{n}k^{2n}}{\left(2n\right)!}M_{0,2n}\left(t\right)\label{eq:InvFourf0}
\end{equation}
To lighten the algebra, we will continue to use $\lambda$ instead
of $k$ for a while. For practical use, we need to simplify the series
entering each moment $M_{0,2n}$, starting from those shown in Eq.(\ref{eq:M2s}).
The higher moments (a few of them can be found in Appendix \ref{sec:AppendA})
contain more terms and quickly become unmanageable without computer
algebra. In the next section, we will sum up the series in Eq.(\ref{eq:InvFourf0})
by extracting the dominant terms from each moment in the sum, depending
on $t$. It is crucial to sum up all the moments with no truncation,
as we mentioned and will further discuss in the next section \footnote{Some of the earlier studies of FP equation also pursued moment approaches,
e.g., \cite{Shalchi2006}. However, a set of moments similar (to $M_{ij}$)
has been truncated typically at $i_{{\rm max}}$ or $j_{\max}$ no
larger than two. Besides, no spatial profile for $f_{0}$ similar
to Eq.(\ref{eq:InvFourf0}) has been given in the above reference,
so we make no comparison with our results. }. 

To emphasise the role of the anisotropic part of particle distribution,
we provide an explicit expression for the moments $M_{0,2n}$ in Eq.(\ref{eq:InvFourf0})
through the lower order, dipolar, moments $M_{1,2n-1}$

\begin{equation}
M_{0,2n}\left(t\right)=2n\int_{0}^{t}M_{1,2n-1}\left(t^{\prime}\right)dt^{\prime}\label{eq:M2nOft}
\end{equation}
\foreignlanguage{american}{The isotropic part of the solution of Eq.(\ref{eq:FPundim})
is thus given by eqs.(\ref{eq:InvFourf0}), (\ref{eq:M2nOft}) and
(\ref{eq:MomSol}).}

\section{Simplified forms of the solution}

Equations (\ref{eq:InvFourf0}) and (\ref{eq:M2nOft}) provide an
exact closed form solution of the Fokker-Planck Eq.(\ref{eq:FPundim}).
The calculation of the moments $M_{0,2n}$ is relatively straightforward.
Using Eq.(\ref{eq:MomSol}) and integrating by parts one obtains $M_{0,2n}$
to any order $n$ in form of polynomials in $t$ and $e^{-t}$: $\sum_{k,l}C_{kl}t^{k}e^{-lt},$
with the constant matrix elements $C_{kl}$ that can also be recursively
obtained from Eq.(\ref{eq:MomSol}). However, the expressions for
$M_{0,2n}$ grow rapidly in length with $n$, and some computer algebra
is virtually required to calculate the series in the Fourier integral
in Eq.(\ref{eq:InvFourf0}). Therefore, for practical use, a simplified
approximation of the series of Fourier integrals is desirable. Before
turning to its derivation, we note that the CR distribution develops
two moving sharp fronts in the profile $f_{0}\left(x,t\right)$, as
discussed in Sec.\ref{subsec:Ballistic-propagation-regime}. Sharp
fronts are dominated by the contributions from $k\gg1$ in the Fourier
integral given by Eq.(\ref{eq:InvFourf0}). This reaffirms our earlier
statement that the series in $n$ should not be truncated at any finite
$n$.

\subsection{Ballistic and transdiffusive phases \label{subsec:Ballistic-and-transdiffusive}}

We begin summing up the infinite series entering the moment generating
function given by Eq.(\ref{eq:MomGenFunc}) for time $t\lesssim1$
relevant for a ballistic and (poorly understood) transdiffusive propagation.
Each term $M_{0,2n}\left(t\right)$ of the series is calculated using
a three-term expansion in powers of $t$ at each power of $\lambda$:

\begin{equation}
\begin{split}f_{\lambda}= & 1+\frac{\lambda^{2}}{2!}\frac{t^{2}}{3}\left(1-\frac{2}{3}t+\frac{t^{2}}{3}+\dots\right)+\\
 & \frac{\lambda^{4}}{4!}\frac{t^{4}}{5}\left(1-\frac{4}{3}t+\frac{58}{45}t^{2}+\dots\right)+\\
 & \frac{\lambda^{6}}{6!}\frac{t^{6}}{7}\left(1-\frac{6}{3}t+\frac{43}{15}t^{2}+\dots\right)+\dots
\end{split}
\label{eq:fLamExp}
\end{equation}
The first two terms in all parenthetical expressions suggest to introduce
the following variable instead of $t$:

\begin{equation}
t^{\prime}=t\left(1-\frac{t}{3}\right).\label{eq:tprimeOft}
\end{equation}
The ``retarded time'' $t^{\prime}$ is related to the second moment
$M_{0,2}\left(t\right)$ as it slows down similarly during the transition
from the ballistic to transdiffusive phase, $M_{0,2}\approx t^{\prime2}/3$,
for $t\lesssim1$. We will use this relation between $t^{\prime}$
and $M_{02}\left(t\right)$ later. Meanwhile, the two leading terms
in $t<1$ at each power of $\lambda$ in Eq.(\ref{eq:fLamExp}) can
be written as powers of $t^{\prime}$ 

\[
\lambda^{n}t^{n}\left(1-\frac{n}{3}t+\dots\right)\approx\lambda^{n}t^{\prime n}
\]
and summed up straightforwardly for $n\to\infty$. The remaining terms
($\sim t^{2}$ in the parentheses in eq.{[}\ref{eq:fLamExp}{]}) can
also be summed up (see Appendix \ref{sec:AppendA}). To the same order
in $t\ll1$, \emph{also valid for large }$\lambda t\gg1$ but still
restricted by $\lambda t^{2}\ll1$, we rewrite the series in Eq.(\ref{eq:fLamExp})
as follows

\begin{equation}
f_{\lambda}\approx\frac{1}{\lambda t^{\prime}}\sinh\left(\lambda t^{\prime}\right)e^{\lambda^{2}\Delta^{2}/4}\label{eq:flamOftPr}
\end{equation}
where

\begin{equation}
\Delta=\begin{cases}
2t^{\prime1/2}t^{3/2}/3\sqrt{3}, & \lambda t\ll1\\
2t^{\prime1/2}t^{3/2}/3\sqrt{5}, & \lambda t\gg1
\end{cases}\label{eq:DeltaCases}
\end{equation}

The full coverage of $f_{\lambda}$ for $0<\lambda t<\infty$ is vital
to a correct description of sharp fronts, implying $\lambda\to\infty$
($k\to\infty$ in Fourier integral). The additional variable $\Delta\left(\lambda,t\right)$
in the above solution changes between its limits at $\lambda t\ll1$
and $\lambda t\gg1$ only insignificantly which offers an opportunity
to use the moment generating function in Eq.(\ref{eq:flamOftPr})
for a full description of the solution $f_{0}\left(x,t\right)$ for
$0<t<\infty$, including the stage when the sharp fronts smear out
(see below). However, the transition between the two limiting cases
is important for such description and discussed further in Sec.\ref{subsec:Unified-Propagator}. 

Using the Fourier spectral parameter $k=i\lambda$ and performing
the inverse Fourier transform according to Eq.(\ref{eq:InvFourf0}),
from Eq.(\ref{eq:flamOftPr}) we obtain

\selectlanguage{american}%
\begin{equation}
f_{0}\left(x,t\right)\approx\frac{1}{4t^{\prime}}\left[{\rm erf}\left(\frac{x+t^{\prime}}{\Delta}\right)-{\rm erf}\left(\frac{x-t^{\prime}}{\Delta}\right)\right]\label{eq:f0OfxtErf}
\end{equation}
Here erf stands for the error function and, again, the essential steps
of the derivation can be found in Appendix \ref{sec:AppendA}. By
construction, this simple result constitutes an approximate, pitch
angle averaged Green's function for Eq.(\ref{eq:FPundim}), which,
as it should, satisfies the condition $f_{0}\to\delta\left(x\right),$
for $t\to0$. As we mentioned and will argue further In Sec.\ref{subsec:Unified-Propagator},
Eq.(\ref{eq:f0OfxtErf}) reaches far beyond its formal validity range,
$t\simeq t^{\prime}<1$, but some conclusions can be drawn immediately
from its present form. 

The spreading of an infinitely narrow initial peak, $f_{0}\left(x,0\right)=\delta\left(x\right)$,
proceeds in the following phases. During a ballistic phase of propagation,
that is strictly at $t\ll1$, the particle distribution $f_{0}\left(x,t\right)$
is best represented by an expanding 'box' of the height $f_{0}=1/2t^{\prime}$
in the region $\left|x\right|<t^{\prime}$. Its edges propagate in
opposite directions along the 'trajectories', $x=\pm t\left(1-t/3\right)=\pm t^{\prime}.$
The box walls are initially much thinner than the box itself, $\Delta\left(t\right)\ll l=2t^{\prime}$,
since $\Delta/l\sim t\ll1$. However, as the expansion progresses
the box walls thicken to its total size $\sim l$ and the process
transitions into a diffusive phase that we consider in the next section.
As the expansion progresses in a retarded time, $t^{\prime}=t-t^{2}/3$,
it slows down. Evidently, this process cannot be followed far enough
in time using Eq.(\ref{eq:f0OfxtErf}), already for the reason that
$t^{\prime}$ becomes negative for $t>3$. Clearly, this problem occurs
from the limitation of the approximate summation of the series in
Eq.(\ref{eq:fLamExp}). As we will show in Sec.\ref{subsec:Unified-Propagator},
Eq.(\ref{eq:f0OfxtErf}) has a potential for much better approximation
of the exact solution at $t\sim1$, if supplemented by more accurate
expressions for $\Delta\left(t\right)$ and $t^{\prime}\left(t\right)$.
To better understand the transdiffusive regime at $t\sim1,$ we first
consider the opposite limit of diffusive particle propagaion at $t\gg1$. 

\subsection{Transition to diffusive propagation\label{subsec:Transition-to-diffusive}}

For $t\gg1$ all terms containing powers of $e^{-t}$ in the expansion
given by Eq.(\ref{eq:MomGenFunc}) can be discarded and only the highest
powers of $t$ need to be retained. Upon extracting such terms from
each $M_{0,2n}\left(t\right)$ by using the solution for the moments
in eqs.(\ref{eq:MomSol}) and (\ref{eq:M2nOft}) (see also the expressions
of the first few moments in Appendix \ref{sec:AppendA}), we may write
the moment generation function in Eq.(\ref{eq:MomGenFunc}) as

\begin{equation}
f_{\lambda}\left(t\right)=\sum_{n=0}^{\infty}\frac{\lambda^{2n}}{\left(2n\right)!}M_{0,2n}\left(t\right)\approx\sum_{n=0}^{\infty}\frac{\left(2n-1\right)!!}{\left(2n\right)!}\lambda^{2n}\left(\frac{t}{3}\right)^{n}\label{eq:fLamLargeT}
\end{equation}
By writing $\left(2n-1\right)!!/\left(2n\right)!=2^{-n}/n!$, the
latter series can be summed up straightforwardly to yield

\begin{equation}
f_{\lambda}=e^{\lambda^{2}t/6}\label{eq:flamGaus}
\end{equation}
After replacing $\lambda=ik$ and performing an inverse Fourier transform
we obtain the conventional diffusive solution

\begin{equation}
f_{0}\left(x,t\right)=\frac{1}{2\pi}\int_{-\infty}^{\infty}dke^{ikx-k^{2}t/6}=\sqrt{\frac{3}{2\pi t}}e^{-3x^{2}/2t}\label{eq:Gauss}
\end{equation}
Now that we have obtained $f_{0}\left(x,t\right)$ for both small
and large $t$, we turn to the most interesting transdiffusive regime
between these two limiting cases. Before doing so, we note that the
last expression for $f_{0}$, valid for $t\gg1$, can be improved
by adding the next order terms to the series in Eq.(\ref{eq:fLamLargeT}).
At the same time, the $t\lesssim1$ and $t\gg1$ results given by
eqs.(\ref{eq:f0OfxtErf}) and (\ref{eq:flamGaus}), respectively,
already allow us to synthesize them into an approximate, uniformly
valid (for all $t$) propagator. We, therefore, consider it in the
next subsection and defer a rigorous matching of the three propagation
regimes to a future study.

\subsection{Unified CR Propagator\label{subsec:Unified-Propagator}}

\selectlanguage{english}%
Although our derivation of the simplified CR propagator in Eq.(\ref{eq:f0OfxtErf})
relies on the condition $t<1$, this condition can be relaxed, provided
that the two time-dependent variables of the propagator, $\Delta\left(t\right)$
and $t^{\prime}\left(t\right)$, are properly redefined. As we will
see, the propagator then accurately describes the solution for all
$t$, including $t\to\infty$. To examine this premise we rewrite
Eq.(\ref{eq:f0OfxtErf}) accordingly

\begin{equation}
f_{0}\left(x,t\right)\approx\frac{1}{4y}\left[{\rm erf}\left(\frac{x+y}{\Delta}\right)-{\rm erf}\left(\frac{x-y}{\Delta}\right)\right].\label{eq:f0Univers}
\end{equation}
So far, we only know $\Delta\left(t\right)$ and $y\left(t\right)\approx t^{\prime}$
at $t\lesssim1$ (see Sec.\ref{subsec:Ballistic-and-transdiffusive}),
so they are left to be determined for $t>1$. As may be seen from
Eq.(\ref{eq:f0Univers}), $\pm y\left(t\right)$ are the coordinates
of two fronts propagating in opposite directions from the initial
$\delta-$ pulse at $x=0$, while $\Delta\left(t\right)$ is the growing
front width. Our hope that Eq.(\ref{eq:f0Univers}) approximates the
true solution better than its prototype in Eq.(\ref{eq:f0OfxtErf}),
is based on the following observations:
\begin{enumerate}
\item It recovers a correct expansion for small $t<1$, both for sharp and
smooth fronts ($\lambda t>1$ and $\lambda t<1$ cases), considered
in Sec. \ref{subsec:Ballistic-and-transdiffusive}
\item It recovers the correct solution for $t\gg1$ in Eq.(\ref{eq:Gauss}).
\item It conserves the total number of particles, for arbitrary $y$ and
$\Delta$
\end{enumerate}
\selectlanguage{american}%
The last statement can be verified by a direct substitution of $f_{0}$
from Eq.(\ref{eq:f0Univers}) 

\[
\begin{split}\int_{-\infty}^{\infty}f_{0}\left(x,t\right)dx=\\
\frac{1}{2\sqrt{\pi}y}\int_{-\infty}^{\infty}dx\left[\int_{0}^{\left(x+y\right)/\Delta}-\int_{0}^{\left(x-y\right)/\Delta}\right]e^{-x^{\prime2}}dx^{\prime}=1 & ,
\end{split}
\]
with the last relation following, e.g., from doing the outer integral
by parts in $x$ and then changing the integration variable in the
both remaining integrals, $x\to x\pm y$. The result (1) was derived
in Sec.\ref{subsec:Ballistic-and-transdiffusive}. The statement (2)
can be verified by expanding the r.h.s. of Eq.(\ref{eq:f0Univers})
in small $y/\Delta\ll1$ which recovers the diffusive solution in
Eq.(\ref{eq:Gauss}) under the condition $\Delta=\sqrt{2t/3}$. 

Based on the above considerations and the results of Secs.\ref{subsec:Ballistic-and-transdiffusive}
and \ref{subsec:Transition-to-diffusive}, we require $\Delta\left(t\right)$
and $y\left(t\right)$ to behave at small and large $t$, respectively,
as follows:

\begin{equation}
\Delta=\begin{cases}
\frac{2}{3\sqrt{5}}t^{2}, & t\ll1,\;{\rm ballistic}\\
\frac{2}{3\sqrt{3}}t^{2}, & t\ll1,\;{\rm transdiffusive}\\
\sqrt{2t/3,} & t\gg1,\;\;{\rm diffusive}
\end{cases}\label{eq:Delta3lines}
\end{equation}

\[
y=\begin{cases}
t^{\prime}, & t\ll1\\
o\left(\sqrt{t}\right), & t\gg1
\end{cases}
\]
The $\Delta$- values for the ballistic and transdiffusive regimes
(both at $t<1)$ differ from each other only by a factor of $\sqrt{5/3}\sim1$.
Furthermore, a trend of changing the $\Delta$ and $y$ time dependencies
from, respectively, $t^{2}$ and $t$ to a slower growth at larger
$t\lesssim1$ has already been rigorously established in Sec.\ref{subsec:Ballistic-and-transdiffusive}
in form of a retarded time, $t\to t\left(1-t/3\right)$. For larger
$t\gtrsim1$ this relation should be improved by summing terms with
higher powers of $t$ in Eq.(\ref{eq:fLamExp}). However, an obvious
similarity of the time dependence of $t^{\prime}\left(t\right)$ and
$\sqrt{3M_{02}\left(t\right)}$ suggests that the moment $M_{02}\left(t\right)$
(and possibly some higher moments) may better describe the quantities
$\Delta\left(t\right)$ and $y(t)$ entering the propagator in Eq.(\ref{eq:f0Univers}).
Indeed, as opposed to $t^{\prime}\left(t\right)$, for example, that
has been consistently derived only for $t<1$, the moments $M_{ij}\left(t\right)$
are calculated exactly for arbitrary $t$. The key observation here
is that the propagator in Eq.(\ref{eq:InvFourf0}) depends on time
not explicitly but only through the moments, $M_{0,2n}\left(t\right)$.
As they all monotonically grow with time, a single moment, such as
$M_{02}\left(t\right)$, suffices to describe the time dependence
of the propagator. On a practical note, all these moments are built
upon the powers of $t$ and $e^{-2t}$, so that they all can be explicitly
expressed through $M_{02}$ and $M_{04}$ from the moment generating
function, given by Eq.(\ref{eq:MomGenFunc}). Another promising avenue
to explore along these lines in a future study is to convert the moment
expansion in Eq.(\ref{eq:MomGenFunc}) into a cumulant expansion.

In this paper, we will use only the second moment $M_{02}\left(t\right)$
to extend our expressions for $\Delta$ and $y$, obtained for $t<1$
and $t\gg1$, over the remainder of the time axis. The particular
choice of $M_{02}$ is supported by its ability to correctly (exactly)
describe the particle spatial dispersion over all the three stages
of their propagation. As we are considering the Green function solution,
we must set $M_{02}\left(0\right)=0$, so the moment $M_{02}$ takes
the following simple form

\selectlanguage{english}%
\begin{equation}
M_{02}=\frac{t}{3}-\frac{1}{6}\left(1-e^{-2t}\right)\label{eq:M02simple}
\end{equation}
\foreignlanguage{american}{Obviously, we cannot recover all the three
cases of $\Delta\left(t\right)$ in Eq.(\ref{eq:DeltaCases}) by making
$t$ in the above formulae simply proportional to $\sqrt{M_{02}}$,
which is correct only for $t<1$. For $t\gg1$, that is in Eq.(\ref{eq:Gauss}),
$t$ should be taken proportional to $M_{02}$ instead. The transition
from ballistic to transdiffusive phase (both within $t<1$) can, in
turn, be calculated by applying the saddle point method to the Fourier
integral in Eq.(\ref{eq:MomGenFunc}). Again, this exercise is planned
for a separate publication. Here we take a practical approach based
on a simple interpolation between different $\Delta\left(t\right)$
and $y\left(t\right)$, and verify the results numerically. So, based
on our $t\ll1$ representation of $\Delta\left(t\right)$ in Eq.(\ref{eq:DeltaCases})
we use the following fit, that should work for not too large $t:$}

\selectlanguage{american}%
\begin{equation}
\Delta=\frac{2}{\sqrt{5}}M_{02}\left(t\right)\left\{ 1+\frac{1}{8}\left[1+\tanh\left(\frac{t-t_{{\rm tr}}}{\Delta t}\right)\right]\right\} \label{eq:DeltaFit}
\end{equation}
Here $t_{{\rm tr}}$ and $\Delta t$ are the transition time and its
duration between ballistic and trans-diffusive phases. The factor
$1/8$ approximately corresponds to the depth of the transition (between
two top lines in Eq.(\ref{eq:Delta3lines})). As for $y\left(t\right)$,
a simple choice consistent with our $t\ll1$ result would be $y=\sqrt{3M_{02}}$.
However, since $y\left(t\right)$ should decay faster than $\Delta\left(t\right)$
for $t\gg1$, we introduce a small correction at $t>t_{tr}^{\prime}$ 

\begin{equation}
y=\sqrt{3M_{02}}\left[1-A\left(t-t_{{\rm tr}}^{\prime}\right)\Theta\left(t-t_{{\rm tr}}^{\prime}\right)\right]\label{eq:yOftFit}
\end{equation}
where $\Theta$ denotese the Heaviside unit function. 

The exact solution of the FP equation is still given in the form of
an infinite series in Eq.(\ref{eq:InvFourf0}), containing terms obtained
recursively. It can hardly be reduced to an expression simple enough
and capable of reproducing the solution with very high accuracy at
the same time. To attain an arbitrary accuracy, the series needs to
be summed up numerically. The approximate propagator in Eq.(\ref{eq:f0Univers})
appears as a plausible alternative in this regard, as it is almost
as simple as the standard Gaussian propagator. To demonstrate that
it is also sufficiently accurate, we plot $f_{0}$ from Eq.(\ref{eq:f0Univers})
with its two input parameters, $\Delta$ and $y$, from eqs.(\ref{eq:DeltaFit})
and (\ref{eq:yOftFit}), respectively. This choice leads to a somewhat
less accurate result for $t\ll1$ than the rigorously obtained $\Delta$
and $y$ in eqs.(\ref{eq:DeltaCases}) and (\ref{eq:f0OfxtErf}),
but we adhere to Eq.(\ref{eq:f0Univers}) because it is accurate for
all $t$.

Shown in Fig.\ref{fig:SixPlots} is the FP solution at different times,
starting from an infinitely narrow initial pulse, $f_{0}\left(x,0\right)=\delta\left(x\right)$.
The solution profiles are plotted using the simplified analytic propagator
from Eq.(\ref{eq:f0Univers}), the standard diffusive solution from
Eq.(\ref{eq:Gauss}), and a direct numerical integration of the FP
equation, as described in Appendix \ref{sec:Numerical-verification-of}.
By its derivation, the analytic propagator is expected to be accurate
for $t<1$, which is successfully confirmed by its comparison with
the numerical results. By contrast, the diffusive solution is particularly
inacurate and acausal at the early times of evolution which was also
not unexpected. Somewhat surprising is that its deviation from the
exact solution continues after quite a few collisions, that is at
$t\gtrsim1$. 

The analytic propagator, on the contrary, demonstrates a very good
agreement with the numerical solution over the total integration time
$t\leq6$ \footnote{Moreover, it successfully reproduces the true FP solution for all
$t$, when an obvious additional constraint $y>0$ is imposed in Eq.(\ref{eq:yOftFit}).
On the other hand, all the three solutions merge into one Gaussian
after $t>5-6$, Fig\ref{fig:Teq6}, so there is no need to discuss
this range. }. It does show a rather expected but surprisingly minor deviation
at $t\sim1$, given that no efforts have yet been made to address
the overlap region between the $t\ll1$ and $t\gg1$ rigorous approximations,
using which the series in the exact solution has been summed up. By
considering the overlap region, the agreement can be systematically
improved which we plan for a future study. The diffusive regime begins
surprisingly late. Even at $t=2-3$, the diffusive solution remains
noticeably acausal.

Surface plots of $f_{0}\left(x,t\right)$ afford additional insight
into the particle propagation. Fig.\ref{fig:Splots} shows them for
the three functions from Fig.\ref{fig:SixPlots}. The surface plots
have an advantage of conveying the information about the time history
of particle intensity observed at a fixed distance from the source.
To obtain this time history, it is sufficient to make a slice of the
surface plot along the plane $x=const.$ It should be remembered,
however, that both $x$ and $t$ variables entering the solution for
$f_{0}\left(x,t\right)$ also depend on particle energy (velocity),
Eq.(\ref{eq:RescaledTimeANDx}). With this in mind, the time histories
obtained for different $x$ may be compared, for example, with the
measurements of solar energetic particles at 1AU for different energies
\cite{Roelof+2002ApJ,Klassen2016A&A}. Propagation of nearly relativistic
electrons is suggested, by these and several other observations, to
be in a scatter-free (ballistic) regime, which is supported by a highly
anisotropic, beam-like, electron distributions. From the standpoint
of the FP solution obtained in this paper, such regime takes place
over times $t\lesssim0.5$ in Fig.\ref{fig:Splots}. The slices along
the $x=const$ planes, that show $f\left(t\right),$ at relatively
small $x$ are characterized by a very steep rise and long decay after
a maximum which is consistent with the observations. However, an inspection
of the anisotropic components of distribution shows that they still
exceed the value of $f_{0}$ near the sharp fronts on the distribution
profile even at times somewhat beyond $t=1$, as may also be inferred
from Fig.\ref{fig:SixPlots}. In this situation, it might be more
appropriate to speak about the transdiffusive rather than ballistic
propagation. Also, anisotropic components of the distribution need
to be extracted from the exact solution for a meaningful comparison
with the observations.
\selectlanguage{english}%

\section{Conclusions}

In this paper, an exact solution for a Fokker-Planck equation, given
in the form of Eq.(\ref{eq:PAscatIntro}), is obtained. The primary
focus has been on the isotropic part of the particle distribution
$f_{0}\left(x,t\right)$ describing an evolution of particle number
density during their propagation from a point source (Green's function).
The propagation is fundamentally simple. It can be categorized into
three phases: ballistic ($t<t_{c})$, transdiffusive ($t\sim t_{c}$)
and diffusive ($t\gg t_{c}$), with $t_{c}$ being the collision time.
In the ballistic phase, the source expands as a ``box'' of size
$\Delta x\propto\sqrt{\left\langle x^{2}\right\rangle }$ with gradually
thickening ``walls.'' The next, transdiffusive phase is marked by
the box's walls thickened to a sizable fraction of the box and its
slower expansion, Figs. \ref{fig:SixPlots} and \ref{fig:Splots}.
Finally, the evolution enters the conventional diffusion phase, Fig.\ref{fig:Teq6}.

The exact closed-form solution of Eq.(\ref{eq:FPundim}) is given
by Eq.(\ref{eq:InvFourf0}) in form of an infinite series in moments
of particle distribution that are, in turn, obtained recursively from
eqs.(\ref{eq:MomSol}) and (\ref{eq:M2nOft}). Eq.(\ref{eq:f0Univers})
and Appendix \ref{sec:AppendA} provide simplified summation formulas
for the series. They adequately describe a point source spreading
through ballistic, transdiffusive and diffusive phases of particle
transport. 

No signatures of a well-known solution to the telegraph equation with
the same initial condition (e.g., \cite{Litvin2016PhPl}) are present
in the exact solution. The signatures are expected in the form of
two sharp peaks attached to the oppositely propagating fronts. Their
absence confirms the earlier conclusion \cite{MS2015} that the telegraph
equation is inconsistent with its parent Fokker-Planck equation except
for a late diffusive phase ($t\gg t_{c}).$ During this phase of particle
propagation, however, a common diffusive reduction of Fokker-Planck
equation, e.g. \cite{Jokipii66}, suffices.

Signatures of superdiffusive propagation regime are also absent in
the exact solution. Such regime is often postulated, e.g., in studies
of diffusive shock acceleration (DSA) \cite{Perrone2013SSRv}, in
the form of a power-law dependence of particle dispersion $\sqrt{\left\langle x^{2}\right\rangle} \propto t^{\alpha}$,
with $1/2<\alpha<1$. According to an exact result shown in Eq.(\ref{eq:M02simple}),
the $\left\langle x^{2}\right\rangle $ time dependence smoothly changes
from the ballistic, $\alpha\to1$, propagation to diffusive one, $\alpha\to1/2$,
without dwelling at any particular value of $\alpha$ inbetween. This
only means, of course, that a simple small-angle scattering model
behind the FP equation does not lead to the superdiffusive transport.
More complicated scattering fields in shock environments, e.g., \cite{MD06},
may result in both superdiffusive (L\'evy flights) and subdiffusive
(long rests) transport anomalies \cite{MetzlerKla2000PhR}. They are,
however, not generic to the DSA (unlike, e.g., Bohm regime) and should
be justified on a case-by-case basis.
\selectlanguage{british}%
\begin{acknowledgments}
I'm indebted to R. Blandford, A. Bykov, and V. Dogiel, along with
other participants at the second San Vito Conference \textquotedblleft Cosmic
Ray Origin beyond the Standard Model,\textquotedblright{} held in
September 2016 in Italy, for interesting discussions of the present
results. A useful exchange of views on the telegraph equation with
Yu. Litvinenko is gratefully acknowledged as well. I also thank two
anonymous referees for furnishing additional references. 

This work was supported by the NASA Astrophysics Theory Program under
Grant No. NNX14AH36G.
\end{acknowledgments}

\selectlanguage{english}%
\appendix

\section{Higher moments of particle distribution and summation formulae \label{sec:AppendA}}

Using Eq.(\ref{eq:MomSol}), after some computer-assisted algebra,
we obtain the following expressions for a few higher moments, $M_{0,2n}$,
needed to compute the full solution given by Eq.(\ref{eq:InvFourf0})

\[
M_{0,4}=\frac{1}{270}e^{-6t}-\frac{t+2}{5}e^{-2t}+\frac{1}{3}t^{2}-\frac{26}{45}t+\frac{107}{270}
\]

\[
\begin{split}M_{0,6}=\frac{1}{31500}e^{-12t}-\frac{3t+2}{1134}e^{-6t}+e^{-2t}\times\\
\left(\frac{3}{10}t^{2}+\frac{639}{350}t+\frac{4743}{1750}\right)+\frac{5}{9}t^{3}-\frac{37}{18}t^{2}+\frac{226}{63}t-\frac{6143}{2268}
\end{split}
\]

\[
\begin{split}M_{0,8}=\frac{1}{6945750}e^{-20t}-\frac{5t+2}{253125}e^{-12t}+\\
\left(\frac{t^{2}}{567}+\frac{11t}{11907}-\frac{59}{27783}\right)e^{-6t}-\\
\left(\frac{14}{25}t^{3}+\frac{858}{125}t^{2}+\frac{151042}{5625}t+\frac{18509371}{506250}\right)e^{-2t}+\\
\frac{35}{27}t^{4}-\frac{224}{27}t^{3}+\frac{3554}{135}t^{2}-\frac{281183}{6075}t+\frac{123403}{3375}
\end{split}
\]
This process can be continued using the recurrence relations in \foreignlanguage{american}{eqs.(\ref{eq:MomSol})
and (\ref{eq:M2nOft}). Our purpose, however, is to derive a simplified,
easy-to-use version of the complete solution. Considering the case
$t<1$, we first expand the moments $M_{0,2n}\left(t\right)$ up to
the order $t^{2n+2}.$ This gives us a series in Eq.(\ref{eq:fLamExp})
for the moment-generating function in which we slightly rearrange
the terms as follows}

\selectlanguage{american}%
\begin{equation}
\begin{split}f_{\lambda}=1+\frac{\lambda^{2}t^{2}}{3!}\left[\left(1-\frac{t}{3}\right)^{2}+\frac{2t^{2}}{9}\dots\right]+\\
\frac{\lambda^{4}t^{4}}{5!}\left[\left(1-\frac{t}{3}\right)^{4}+\frac{28}{45}t^{2}+\dots\right]+\\
\frac{\lambda^{6}t^{6}}{7!}\left[\left(1-\frac{t}{3}\right)^{6}+\frac{6}{5}t^{2}+\dots\right]+\dots
\end{split}
\label{eq:flamApp}
\end{equation}
This representation of the series suggests using a retarded time $t^{\prime}=t\left(1-t/3\right)$
instead of $t$. Separating then the terms in the brackets to form
two individual series, of which the first one can be summed up immediately,
we obtain

\begin{equation}
f_{\lambda}=\frac{1}{2\lambda t^{\prime}}\left(e^{\lambda t^{\prime}}-e^{-\lambda t^{\prime}}\right)+\frac{t^{2}}{45}S\left(\lambda t\right)\label{eq:flam2}
\end{equation}
The remaining series $S\left(\lambda t\right)$ can be represented
in the following compact form

\[
S\left(y\right)\equiv\sum_{n=1}^{\infty}\frac{2n\left(2n+3\right)}{\left(2n+1\right)!}y^{2n}
\]
and summed up by rewriting it as

\[
S=y\frac{d}{dy}\frac{1}{y^{2}}\frac{d}{dy}\sum_{n=1}^{\infty}\frac{\left(y\right)^{2n+3}}{\left(2n+1\right)!}=y\frac{d}{dy}\frac{1}{y^{2}}\frac{d}{dy}y^{2}\left(\sinh y-y\right)
\]
Using this formula, we evaluate Eq.(\ref{eq:flam2}) to its final
form

\begin{equation}
\begin{split}f_{\lambda}\left(t\right)=\frac{1}{\lambda t^{\prime}}\sinh\left(\lambda t^{\prime}\right)+\\
\frac{t^{2}}{45}\left[2\cosh\left(\lambda t\right)+\left(\lambda t-\frac{2}{\lambda t}\right)\sinh\left(\lambda t\right)\right]
\end{split}
\label{eq:flam3}
\end{equation}
It is important that $\lambda t$ can be arbitrarily large here, $\lambda t\gg1$,
even though our expansion technically requires $t<1$. As we emphasized
earlier, large values of $\lambda$ are associated with the contribution
of higher moments in the series in Eq.(\ref{eq:flamApp}) which, in
turn, is responsible for sharp fronts.

The leading term in Eq.(\ref{eq:flam3}) is the first one, while from
the second one we extract a contribution proportional to $\lambda t$.
First, we assume $\lambda t\gg1$ which applies to sharp fronts and
narrow initial CR distributions. Although the ballistic phase precedes
the transdiffusive one, the value of $\lambda t$ is effectively larger
namely during the ballistic phase. It decreases while the fronts become
smoother with increasing time but with the condition $t<1$ held up. 

To the same order of expansion in $t<1$, Eq.(\ref{eq:flam3}) can
then be written as ($\lambda t\gg1$)

\begin{equation}
\begin{split}f_{\lambda}\left(t\right)\approx\frac{1}{\lambda t^{\prime}}\sinh\left(\lambda t^{\prime}\right)\left(1+\frac{\lambda^{2}t^{3}t^{\prime}}{45}\right)\approx\\
\frac{1}{\lambda t^{\prime}}\sinh\left(\lambda t^{\prime}\right)e^{\lambda^{2}t^{3}t^{\prime}/45}
\end{split}
\label{eq:flamWithExp}
\end{equation}
Turning to the opposite case, $\lambda t<1$, instead of the last
expression we have

\begin{equation}
f_{\lambda}\left(t\right)\approx\frac{1}{\lambda t^{\prime}}\sinh\left(\lambda t^{\prime}\right)e^{\lambda^{2}t^{3}t^{\prime}/27}\label{eq:flamExp2}
\end{equation}
Because of the approximation $t^{\prime}\approx t<1$, the product
of $t^{3}$ and $t^{\prime}$ in these formulae can be replaced by
a different combination $t^{m}t^{\prime4-m}$, as it enters a correction
term. From the standpoint of asymptotic expansion in use, there is
no difference between $t$ and $t^{\prime}$ in the highest order
terms included. We used this aspect in obtaining the series expansion
in Eq.(\ref{eq:flamExp2}). It is a remarkable feature of the above
expansion of $f_{\lambda}\left(t\right)$ that it has a very similar
form for small and large values of $\lambda t$. Even more interesting,
perhaps, is that the solution $f_{0}\left(x,t\right)$, obtained below
by inverting the Fourier integral from the above relations, can be
written using the same functional dependence of the solution on the
relevant time-dependent variables in all three phases, Eq.(\ref{eq:f0Univers}).

Replacing $\lambda$ with $-ik$ and substituting Eq.(\ref{eq:flamWithExp})
into the inverse Fourier integral in Eq.(\ref{eq:InvFourf0}), we
obtain

\[
\begin{split}f_{0}\left(x,t\right)=\frac{1}{2\pi}\int e^{ikx}f_{-ik}\left(t\right)dk=\\
\frac{1}{4\pi t^{\prime}}\int_{-\infty}^{x}dx\int_{-\infty}^{\infty}dk\left(e^{ikx_{-}}-e^{-ikx_{+}}\right)e^{-k^{2}t^{3}t^{\prime}/45}
\end{split}
\]
where $x_{\pm}=x\mp t^{\prime}$. By performing a straightforward
integration the result given in Eq.(\ref{eq:f0OfxtErf}) immediately
follows. Note that the last relation is obtained for the case $\lambda t\gg1$,
while in the opposite case Eq.(\ref{eq:flamExp2}) should be used
which simply results in the replacement $45\to27$ in the exponent. 

\section{Numerical verification of the analysis\label{sec:Numerical-verification-of}}

\begin{figure}
\selectlanguage{english}%
\includegraphics[bb=100bp 0bp 792bp 612bp,scale=0.3]{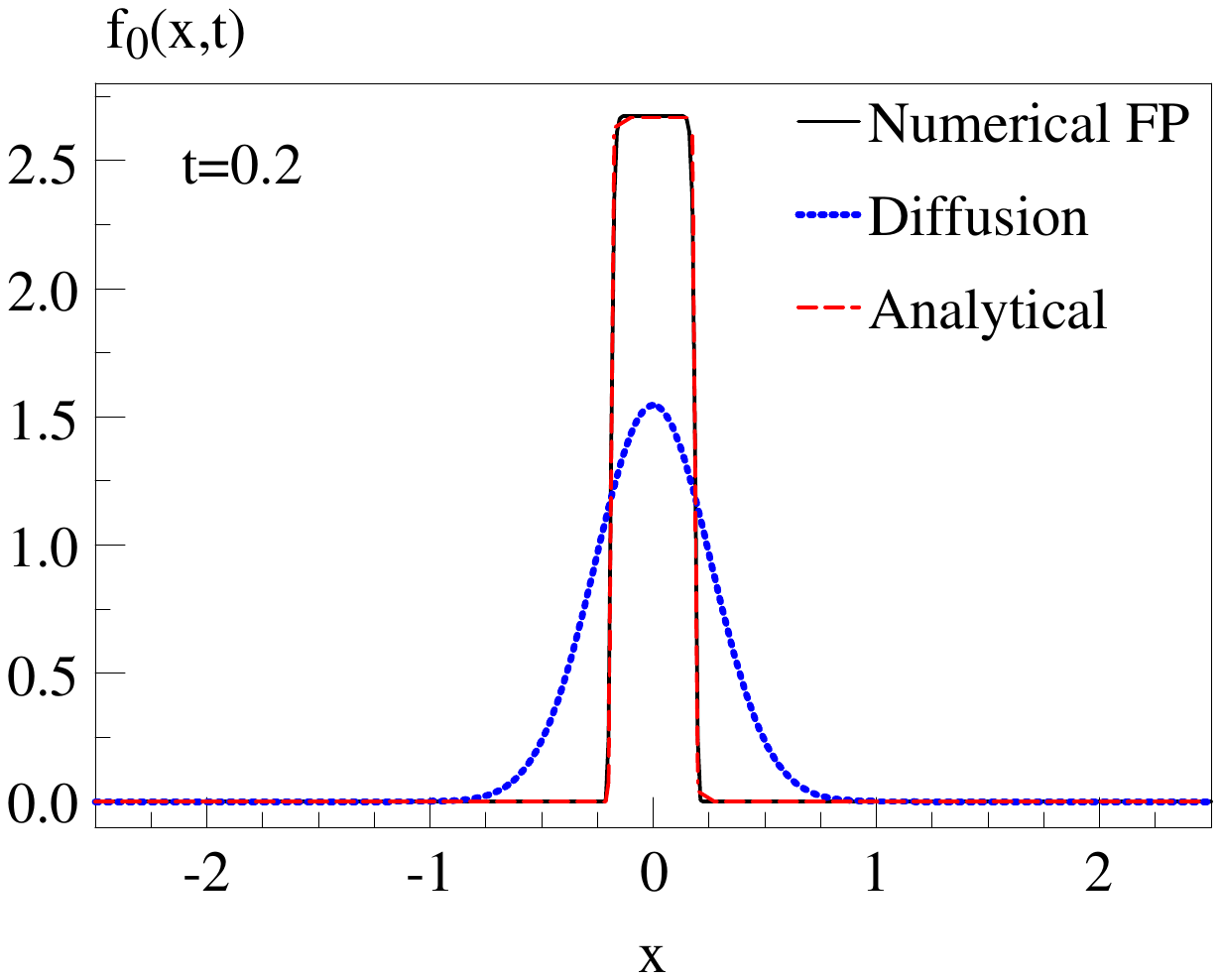}\includegraphics[bb=100bp 0bp 792bp 612bp,scale=0.3]{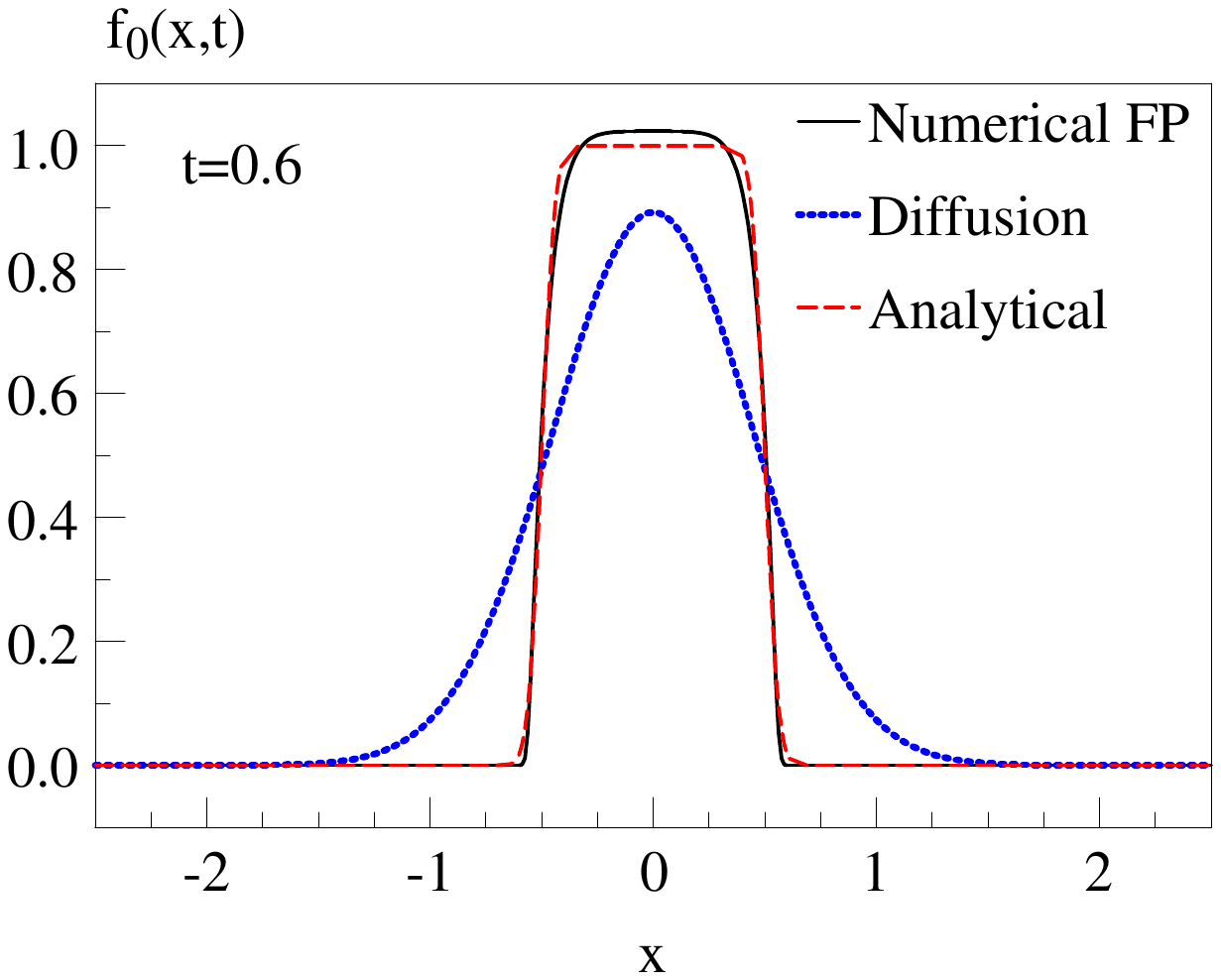}

\includegraphics[bb=100bp 100bp 792bp 512bp,scale=0.3]{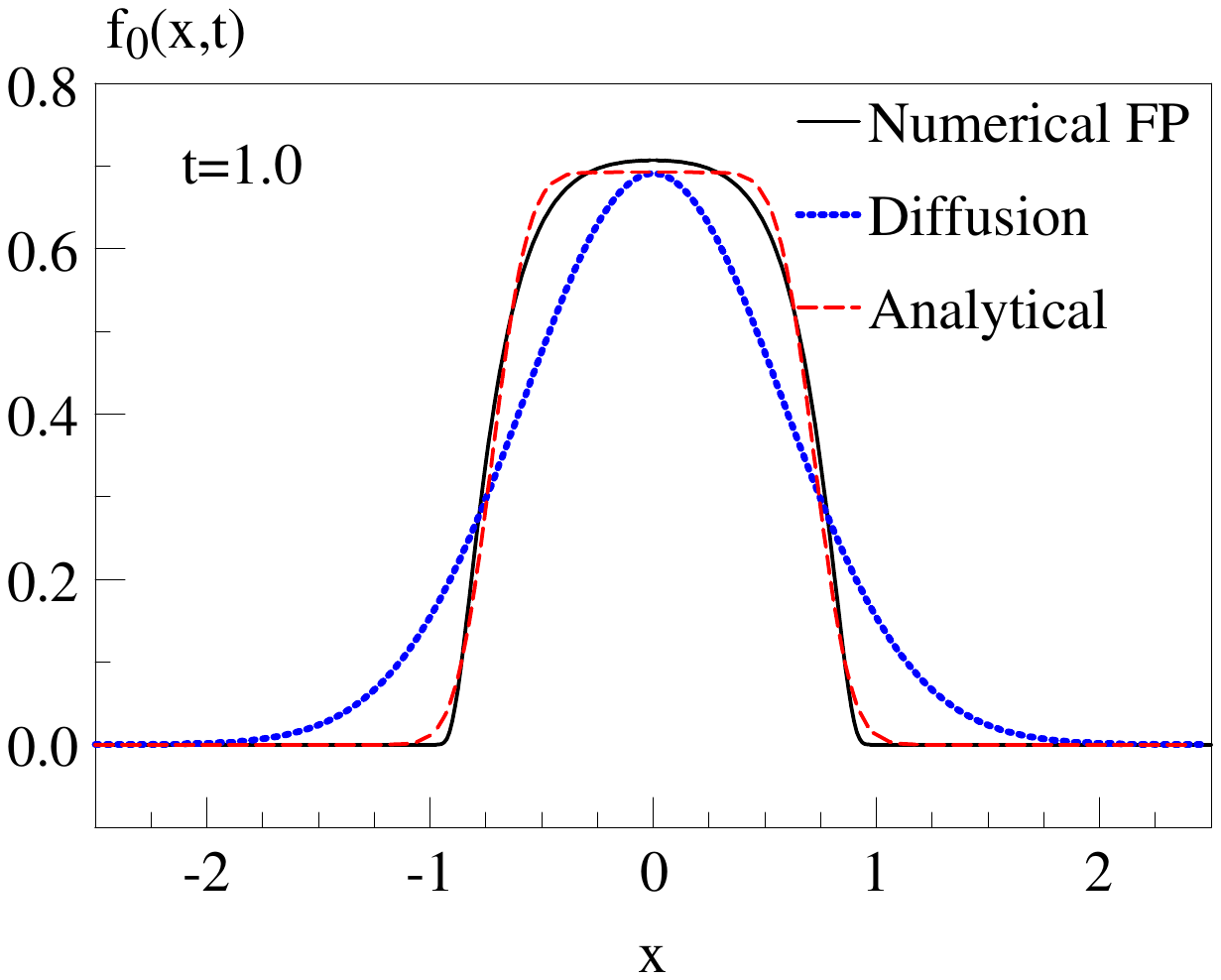}\includegraphics[bb=100bp 100bp 792bp 512bp,scale=0.3]{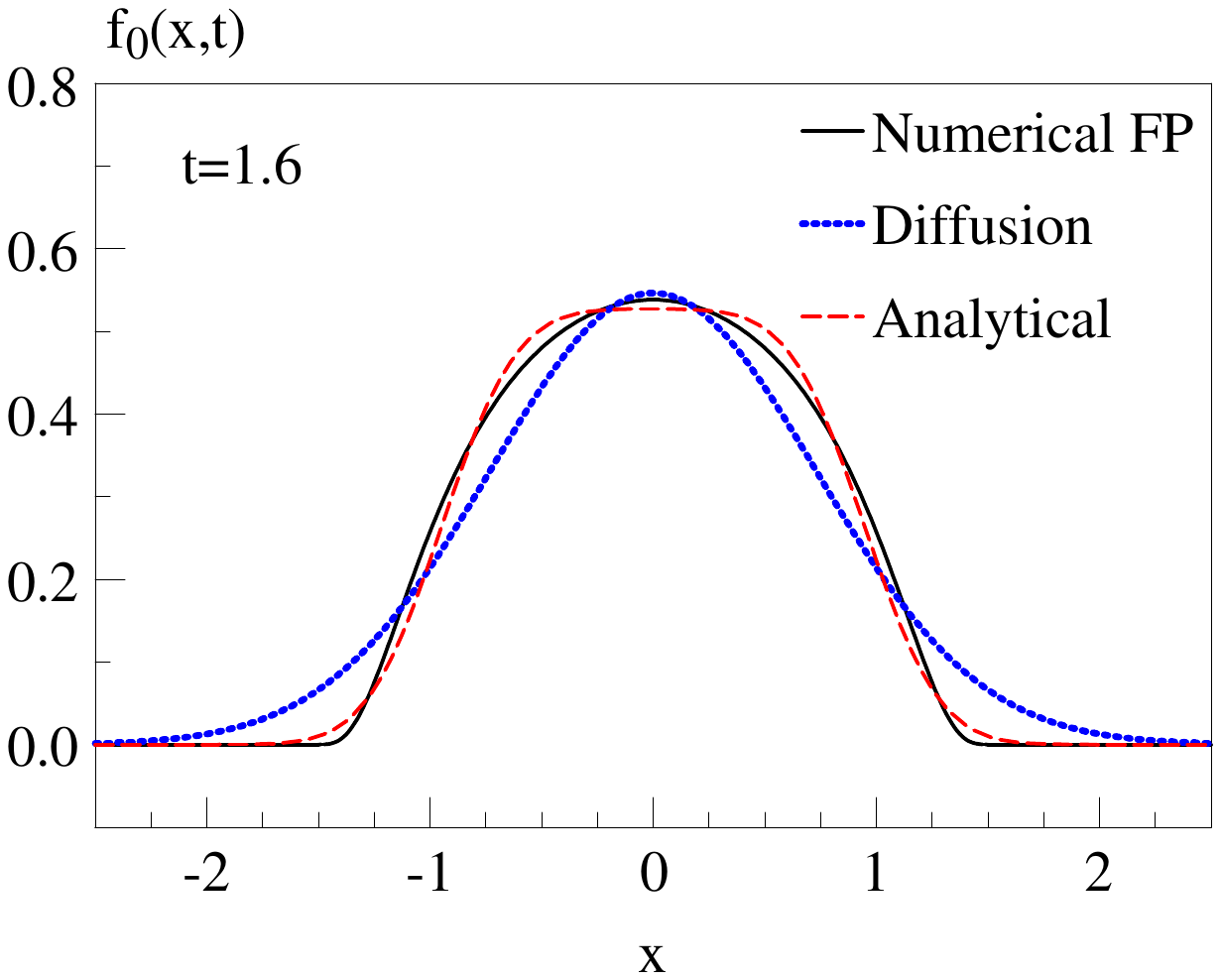}

\includegraphics[bb=100bp 100bp 792bp 612bp,scale=0.3]{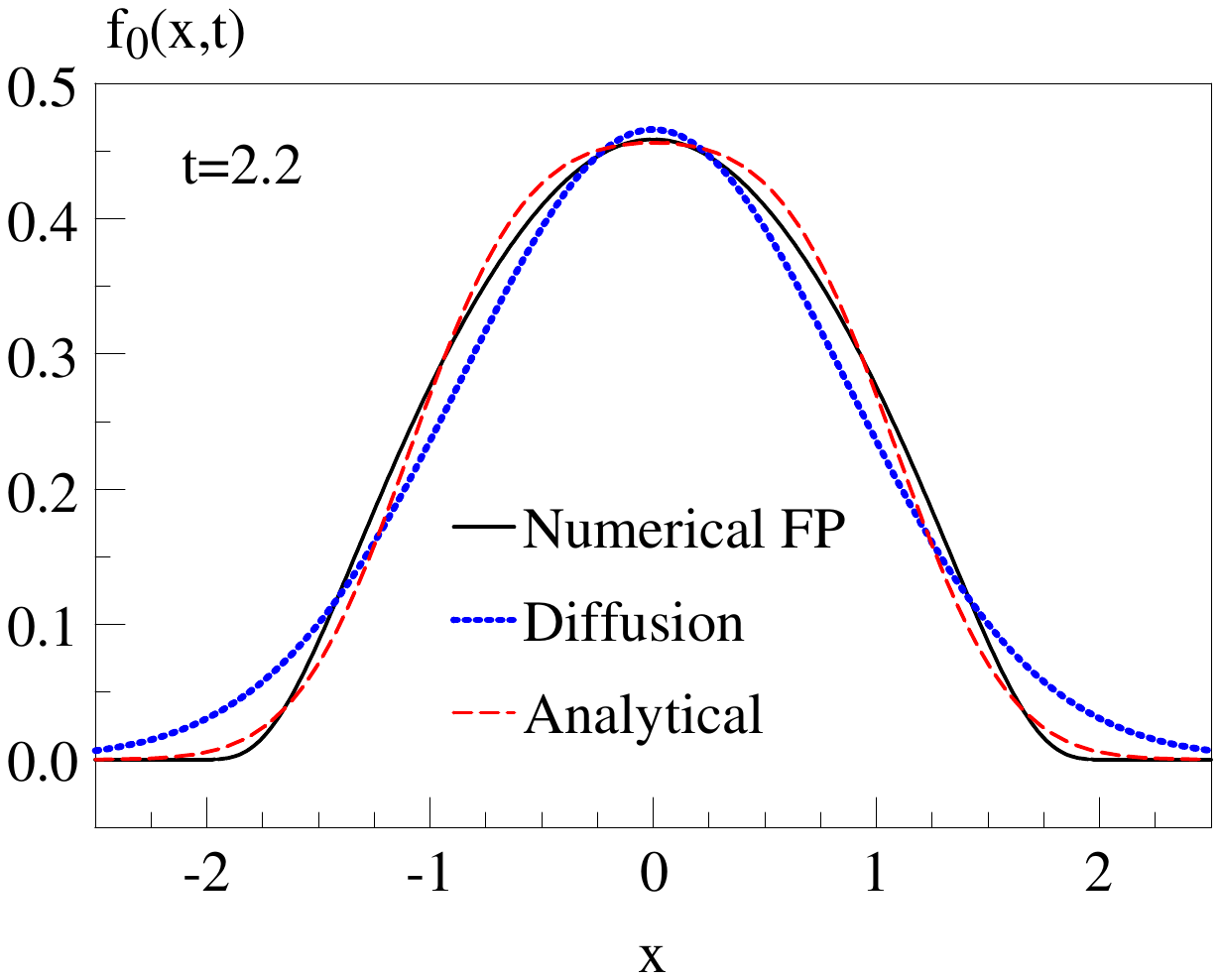}\includegraphics[bb=100bp 100bp 792bp 612bp,scale=0.3]{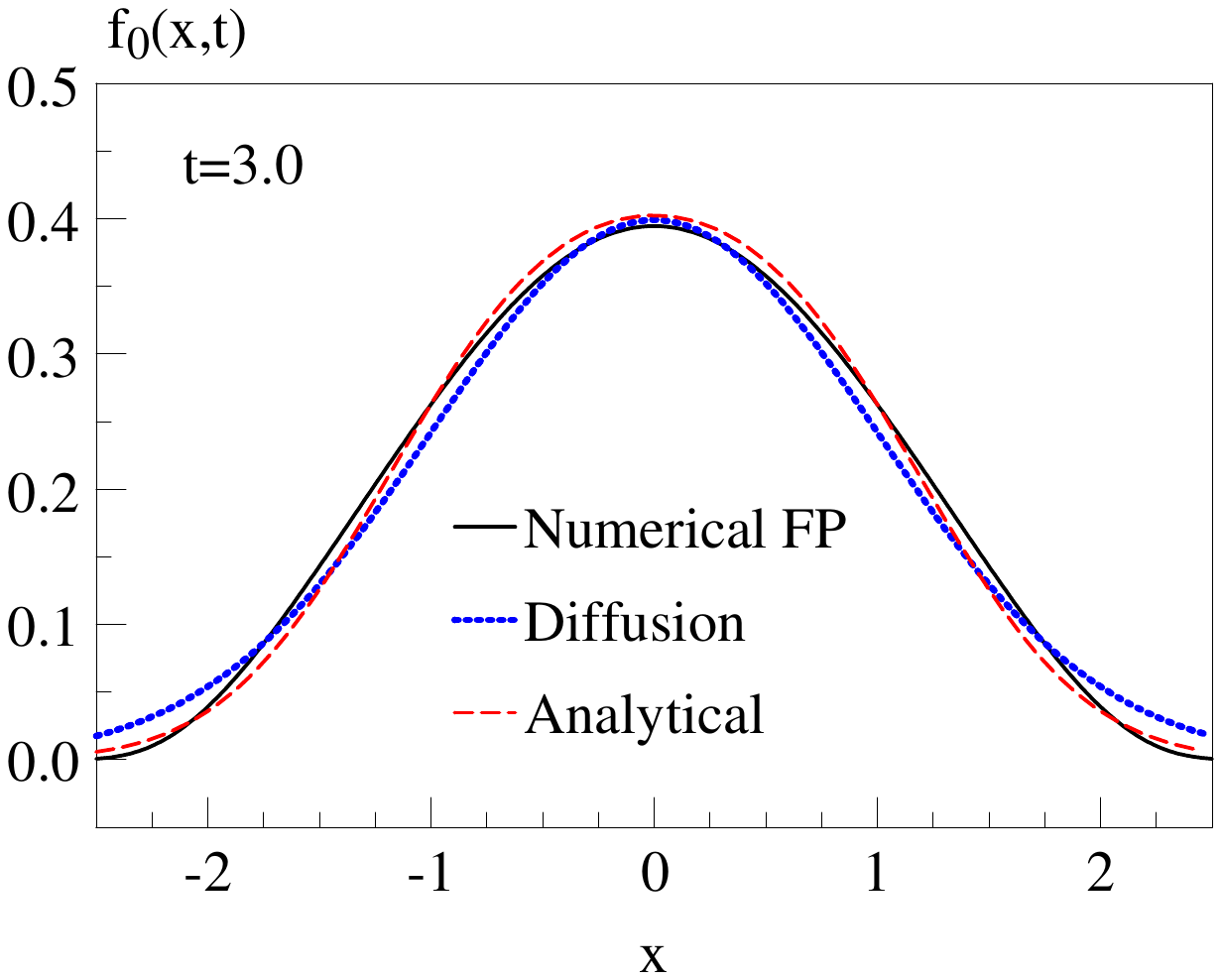}

\selectlanguage{american}%
\caption{Fundamental solution of the Fokker-Planck equation shown for its isotropic
component, $f_{0}\left(x,t\right)=\left\langle f\left(x,\mu,t\right)\right\rangle $
at six different times. Analytic approximation is taken from Eq.(\ref{eq:f0Univers})
with $\Delta$ and $y$ from eqs.(\ref{eq:DeltaFit}) and (\ref{eq:yOftFit}),
respectively. The numerical parameters used in these formulas are,
respectively, $\Delta t=0.2$, $t_{tr}=0.85$, $A\approx0.1$, $t_{tr}^{\prime}=0.6.$
The diffusive (Gaussian) solution, that is shown for comparison, is
obtained from Eq.(\ref{eq:Gauss}). The numerical solution is described
in Appendix \ref{sec:Numerical-verification-of}. \label{fig:SixPlots}}
\end{figure}

\selectlanguage{english}%
Regarding a full expansion of particle distribution in Legendre polynomials

\begin{equation}
f\left(x,\mu,t\right)=\sum_{n=0}^{\infty}f_{n}\left(x,t\right)P_{n}\left(\mu\right),\label{eq:LegExp}
\end{equation}
so far we have focused on $f_{0}$ as it is an isotropic and only
constituent of the particle distribution that contributes to their
density. This does not mean, of course, that $f_{n}$ with $n>0$
are unimportant. During the ballistic and transdiffusive propagation
phases, a dozen of the first $f_{n}$ are generally of the order of,
or even larger than, $f_{0},$ especially near the two sharp fronts
in Fig.\ref{fig:SixPlots}.

Substituting the above expansion into the FP equation (\ref{eq:FPundim})
one obtains the following system for $f_{n}\left(x,t\right)$

\begin{equation}
\begin{split}\frac{\partial f_{n}}{\partial t}=-n\left(n+1\right)f_{n}-\frac{n}{2n-1}\frac{\partial f_{n-1}}{\partial x}-\\
\frac{n+1}{2n+3}\frac{\partial f_{n+1}}{\partial x}+\epsilon\frac{\partial f_{n}^{2}}{\partial x^{2}}
\end{split}
\label{eq:fnSystem}
\end{equation}
which we solve numerically for $0\le n\le n_{{\rm max}},$ under the
following conditions

\[
f_{n}\equiv0,\;\;\;{\rm for}\;\;\;n<0,\;\;n\ge n_{{\rm max}}
\]

\[
f_{0}\left(x,0\right)=\frac{1}{\sqrt{\pi}\delta}e^{-x^{2}/\delta^{2}},\;\;\;f_{n}\left(x,0\right)=0,\;\;\;n>0
\]

\begin{figure}
\includegraphics[bb=100bp 0bp 792bp 612bp,scale=0.3]{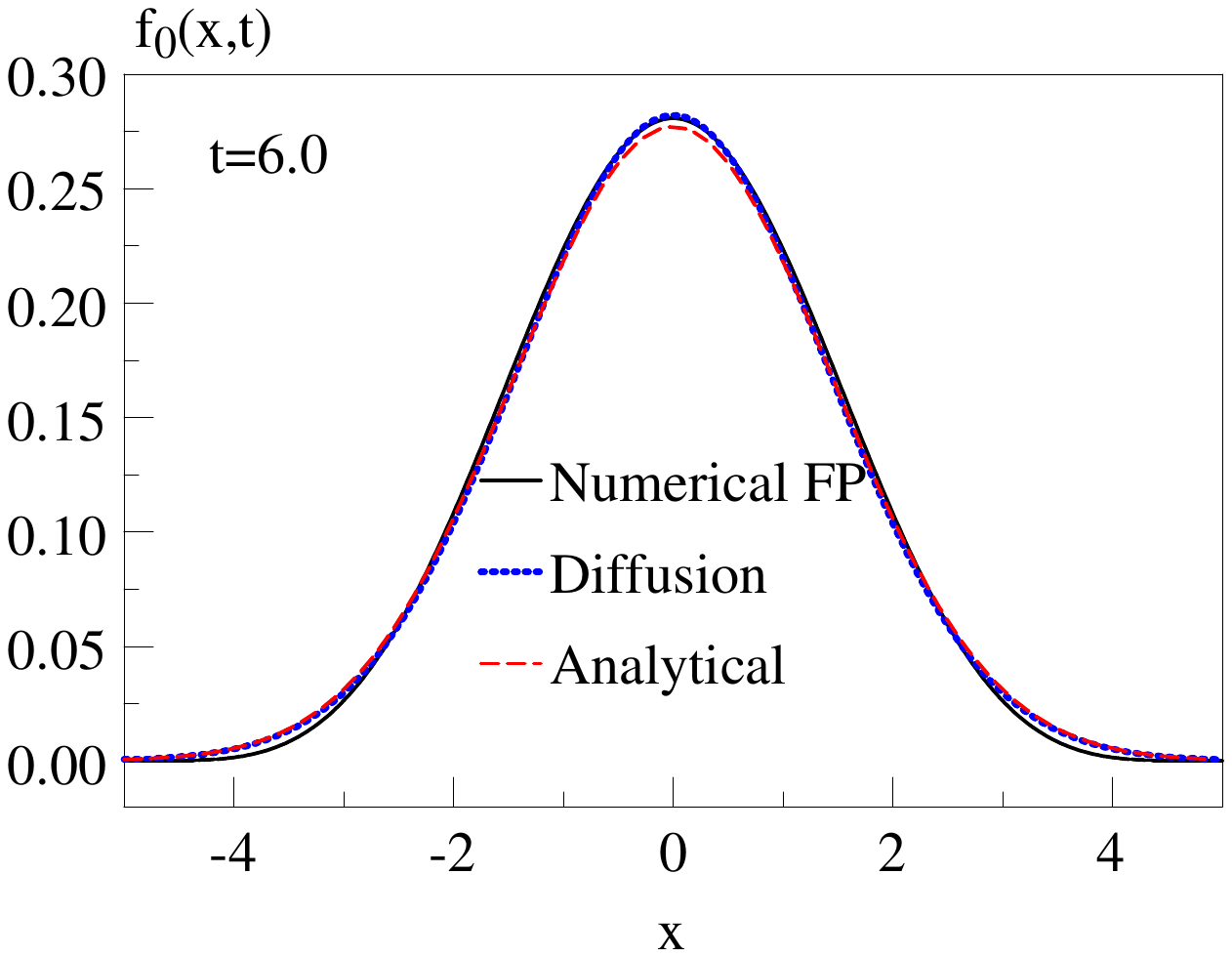}

\caption{Same as in Fig.\ref{fig:SixPlots} but for $t=6.0$\label{fig:Teq6}}
\end{figure}

\begin{figure}
\includegraphics[bb=0bp 50bp 1161bp 670bp,scale=0.25]{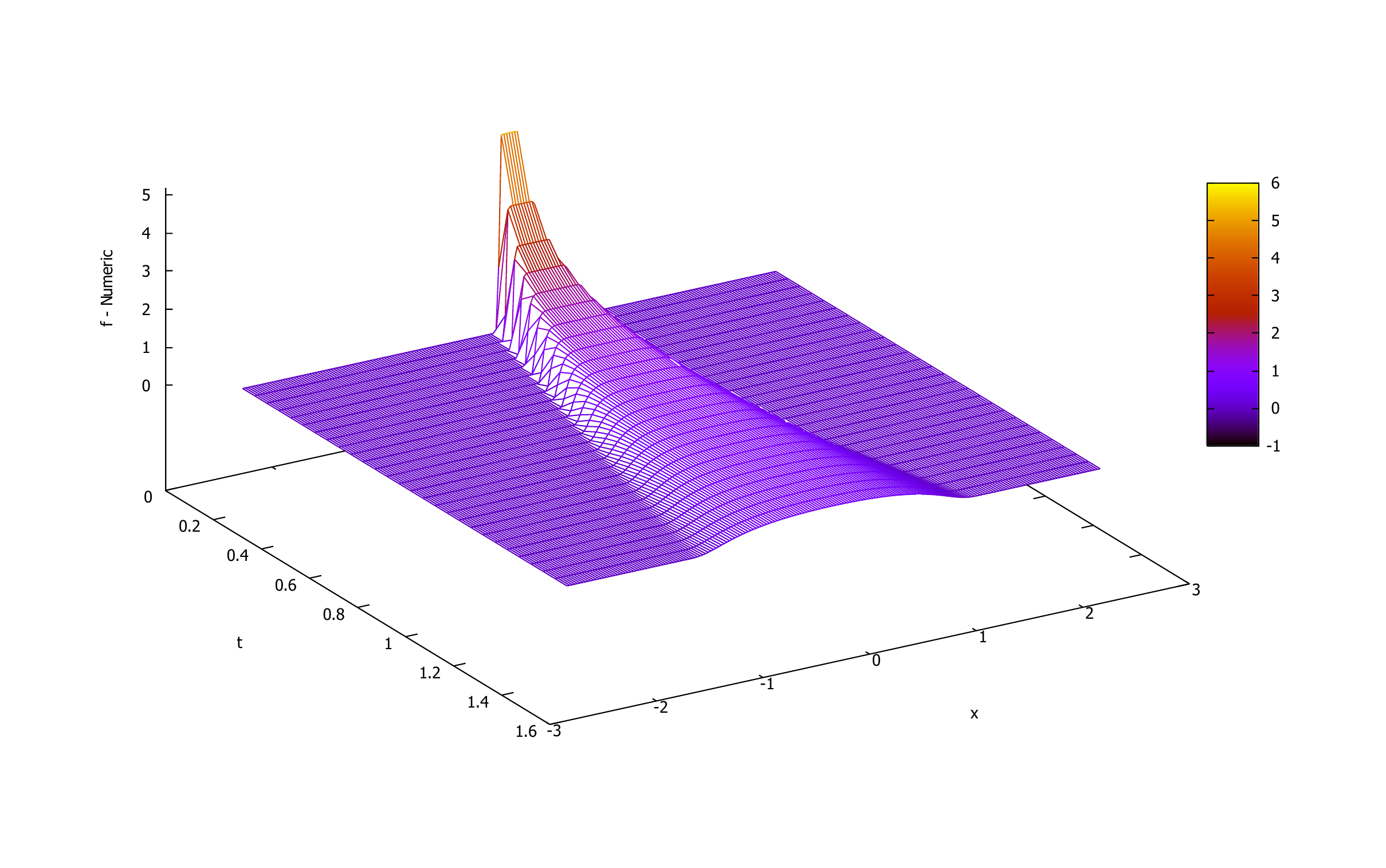}

\includegraphics[bb=0bp 0bp 1161bp 670bp,scale=0.25]{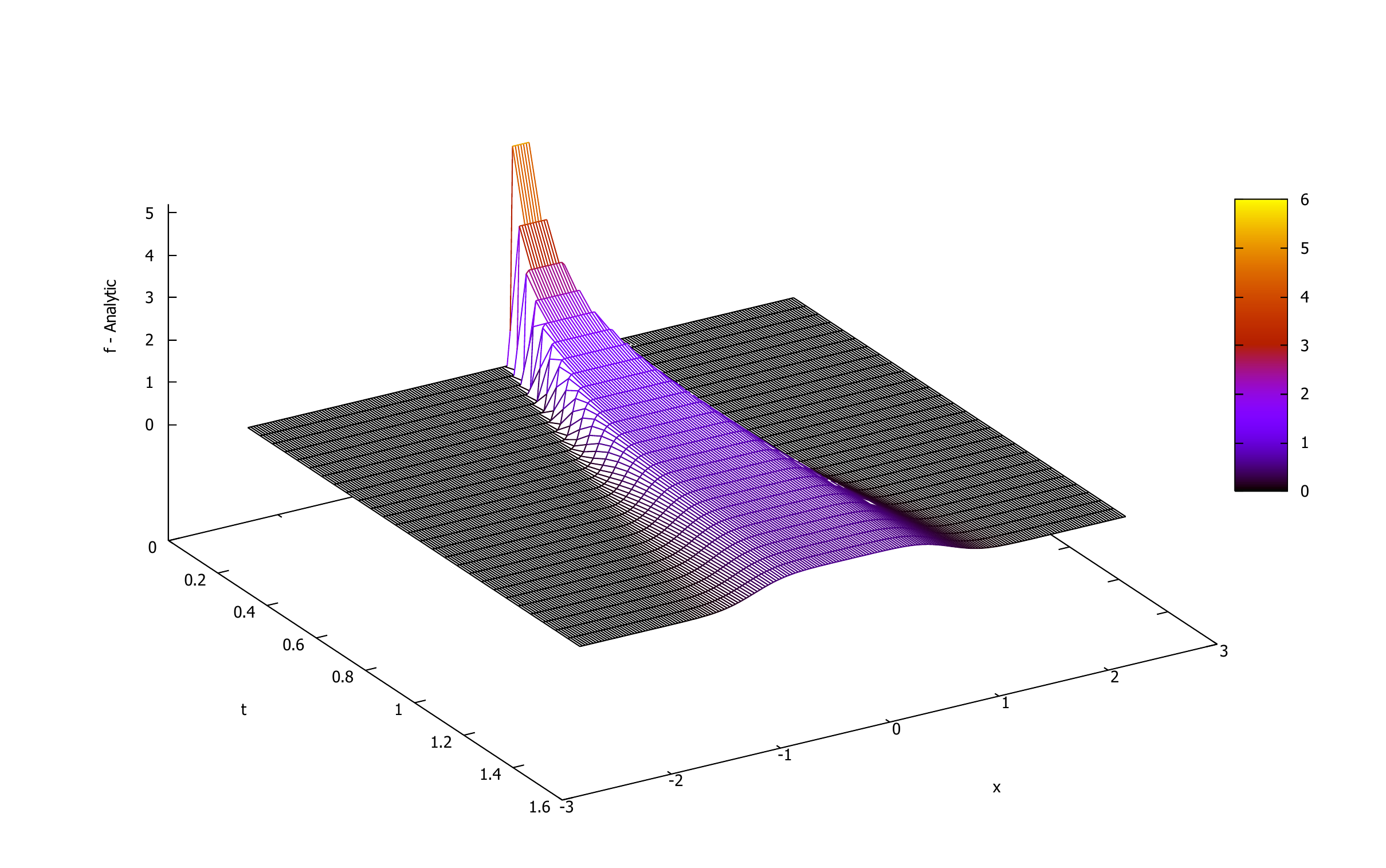}

\includegraphics[bb=0bp 0bp 1161bp 680bp,scale=0.25]{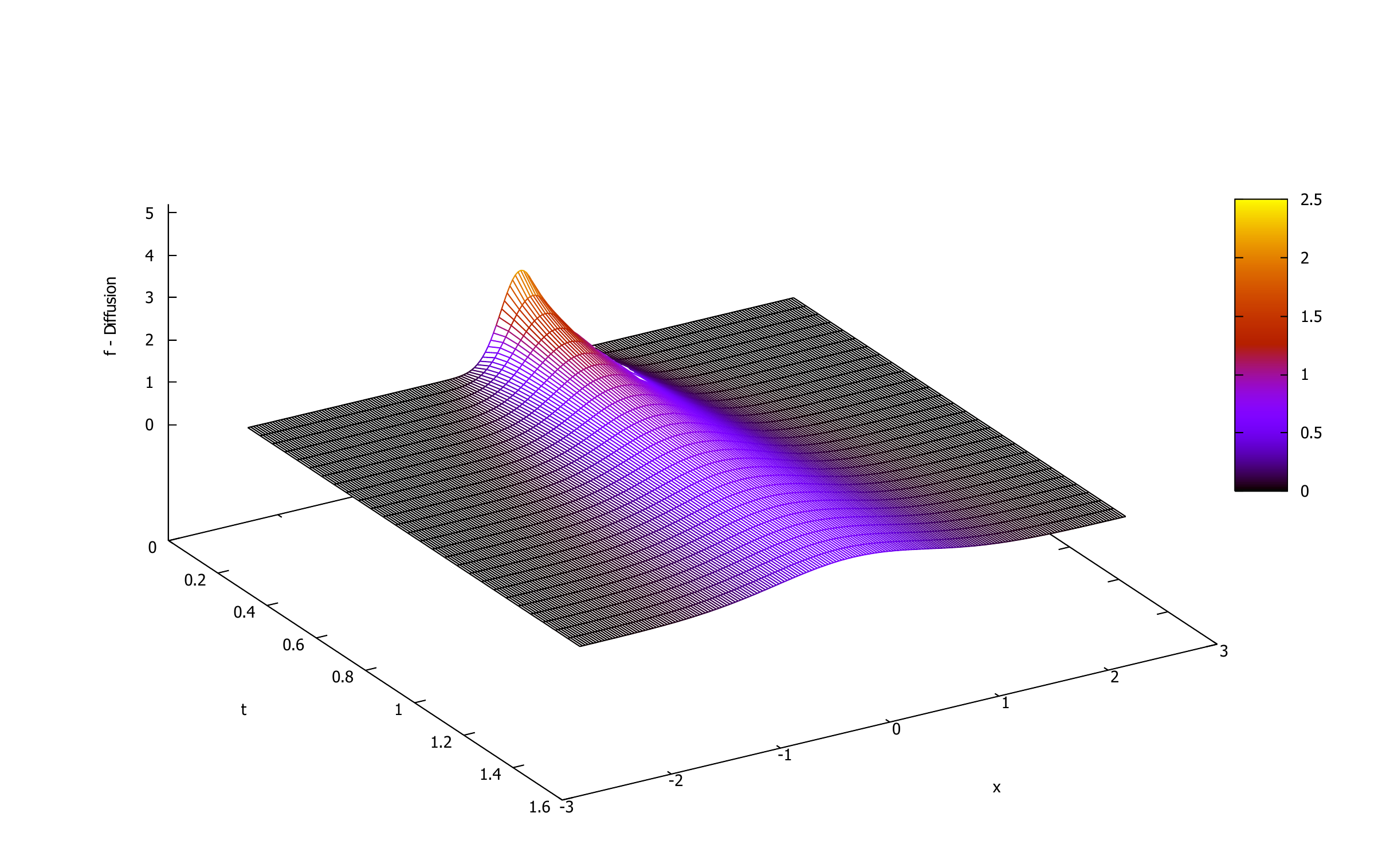}

\caption{\foreignlanguage{american}{The same three solutions (numerical, analytical, and diffusive) as
in Fig.\ref{fig:SixPlots} but shown as surface plots of $f_{0}\left(x,t\right)$
between $t=0.1-1.5$.\label{fig:Splots}}}
\end{figure}

A small diffusive term is added to the r.h.s for wellposedness. Keeping
$\epsilon\lesssim10^{-7}$ makes the integration results practically
insensitive to this parameter, even for rather steep initial conditions
(small $\delta$). The optimum truncation parameter $n_{{\rm max}}$
also depends on the scale of initial condition, for which we take
a Gaussian with the width $\delta\lesssim10^{-3}$ for $f_{0}\left(x,0\right)$
and zero for $n\neq0$. This setting is consistent with a Green function
solution sought for $f_{0}$. Under these conditions, the optimum
$n_{{\rm max}}\simeq16$, beyond which the integration results do
not change noticeably. The integration domain in $-a<x<a$ is taken
large enough to ensure the conditions $f_{n}\left(\pm a,t\right)\ll f_{{\rm 0}}\left(0,t\right)$
(at the maximum of $f_{0}$). The choice of $a$ depends, of course,
on the integration time, as seen from Figs.\ref{fig:SixPlots} and
\ref{fig:Teq6}.

\selectlanguage{american}%
The system in Eq.(\ref{eq:fnSystem}) was integrated numerically for
$f_{n}\left(x,t\right)$, $n_{{\rm max}}=16$ in Eq.(\ref{eq:LegExp})
using an adaptive mesh refinement (collocation) algorithm, described
in, e.g., \cite{MuirBacoli2004}. It is especially well suited for
evolving sharp fronts that are formed during the ballistic and transdiffusive
particle propagation regimes. 

\selectlanguage{english}%
\bibliographystyle{prsty}
\bibliography{FPsol.bbl}

\begin{thebibliography}{10}

\bibitem{Jok1971RvGSP}
J.~R. {Jokipii}, Reviews of Geophysics and Space Physics {\bf 9},  27  (1971).

\bibitem{Voelk73}
H.~J. {V{\"o}lk}, \apss {\bf 25},  471  (1973).

\bibitem{M_PoP2015}
M.~A. {Malkov}, Physics of Plasmas {\bf 22},  091505  (2015).

\bibitem{GrenierRev2015}
I.~A. {Grenier}, J.~H. {Black}, and A.~W. {Strong}, \araa {\bf 53},  199
  (2015).

\bibitem{Jokipii66}
J.~R. {Jokipii}, \apj {\bf 146},  480  (1966).

\bibitem{AharDV94}
F.~A. {Aharonian}, L.~O. {Drury}, and H.~J. {Voelk}, \aap {\bf 285},  645
  (1994).

\bibitem{Note1}
Note, however, that if the pressure of released particles is of the order or
  larger than the magnetic pressure of the ambient medium, the particles are
  self-confined by scattering off self-generated MHD waves. This kind of
  problems should be treated differently, [38-41] from what is described
  further in this paper, as the scattering frequency $D$ would strongly depend
  on $\nabla f$, whereas here we consider the case $D=D\left (v\right )$.

\bibitem{MS2015}
M.~A. {Malkov} and R.~Z. {Sagdeev}, \apj {\bf 808},  157  (2015).

\bibitem{SchwadronTelegraph94}
N.~A. {Schwadron} and T.~I. {Gombosi}, \jgr {\bf 99},  19301  (1994).

\bibitem{Berezinsky2007}
V. {Berezinsky} and A.~Z. {Gazizov}, \apj {\bf 669},  684  (2007).

\bibitem{AloisBerezSuperLum09}
R. {Aloisio}, V. {Berezinsky}, and A. {Gazizov}, \apj {\bf 693},  1275  (2009).

\bibitem{AxfordTelegr65}
W.~I. {Axford}, \planss {\bf 13},  1301  (1965).

\bibitem{GOLDSTEIN01011951}
S. Goldstein, The Quarterly Journal of Mechanics and Applied Mathematics {\bf
  4},  129  (1951).

\bibitem{Earl1974}
J.~A. {Earl}, \apj {\bf 188},  379  (1974).

\bibitem{LitvinEffenSchlick15}
Y.~E. {Litvinenko}, F. {Effenberger}, and R. {Schlickeiser}, \apj {\bf 806},
  217  (2015).

\bibitem{Litvin2016PhPl}
Y.~E. {Litvinenko} and P.~L. {Noble}, Physics of Plasmas {\bf 23},  062901
  (2016).

\bibitem{Note2}
Evidently, the above arguments do not apply to a telegraph equation derived
  from a Boltzmann equation with a simplified scattering operator of the form
  $St\left (f\right )=\left (f_{0}-f\right )/\tau $ (so called BGK, or $\tau -$
  approximation). In this context, the telegraph equation has been studied in,
  e.g., [42-46].

\bibitem{Taylor01011922}
G.~I. Taylor, Proceedings of the London Mathematical Society {\bf s2-20},  196
  (1922).

\bibitem{VVSQL62}
A.~A. Vedenov, E.~P. Velikhov, and R.~Z. Sagdeev, NUCLEAR FUSION, Suppl.2,  465  (1962).

\bibitem{Kennel66}
C.~F. {Kennel} and F. {Engelmann}, Physics of Fluids {\bf 9},  2377  (1966).

\bibitem{Skill75a}
J. {Skilling}, \mnras {\bf 172},  557  (1975).

\bibitem{BlandEich87}
R. {Blandford} and D. {Eichler}, \physrep {\bf 154},  1  (1987).

\bibitem{goldr97}
P. {Goldreich} and S. {Sridhar}, \apj {\bf 485},  680  (1997).

\bibitem{ChandranGS00PhRvL}
B.~D.~G. {Chandran}, Physical Review Letters {\bf 85},  4656  (2000).

\bibitem{Acht81a}
A. {Achterberg}, \aap {\bf 98},  161  (1981).

\bibitem{Milagro10}
M.~A. {Malkov}, P.~H. {Diamond}, L. {O'C.~Drury}, and R.~Z. {Sagdeev}, \apj
  {\bf 721},  750  (2010).

\bibitem{Bykov2014ApJ}
A.~M. {Bykov}, D.~C. {Ellison}, S.~M. {Osipov}, and A.~E. {Vladimirov}, \apj
  {\bf 789},  137  (2014).

\bibitem{Shalchi2009}
A. {Shalchi}, T. {Skoda}, R.~C. {Tautz}, and R. {Schlickeiser}, \aap {\bf 507},
   589  (2009).

\bibitem{reed1975methods}
M. Reed and B. Simon, {\em Methods of modern mathematical physics, Vol. II}
  (Academic Press,~New~York, 1975).

\bibitem{Note3}
Some of the earlier studies of FP equation also pursued moment approaches,
  e.g., [47]. However, a set of moments similar (to $M_{ij}$)
  has been truncated typically at $i_{{\protect \rm max}}$ or $j_{\protect
  \qopname \relax m{max}}$ no larger than two. Besides, no spatial profile for
  $f_{0}$ similar to Eq.(\ref {eq:InvFourf0}) has been given in the above
  reference, so we make no comparison with our results.

\bibitem{Note4}
Moreover, it successfully reproduces the true FP solution for all $t$, when an
  obvious additional constraint $y>0$ is imposed in Eq.(\ref {eq:yOftFit}). On
  the other hand, all the three solutions merge into one Gaussian after
  $t>5-6$, Fig\ref {fig:Teq6}, so there is no need to discuss this range.

\bibitem{Roelof+2002ApJ}
G.~M. {Simnett}, E.~C. {Roelof}, and D.~K. {Haggerty}, \apj {\bf 579},  854
  (2002).

\bibitem{Klassen2016A&A}
A. {Klassen} {\it et~al.}, \aap {\bf 593},  A31  (2016).

\bibitem{Perrone2013SSRv}
D. {Perrone} {\it et~al.}, \ssr {\bf 178},  233  (2013).

\bibitem{MD06}
M.~A. {Malkov} and P.~H. {Diamond}, \apj {\bf 642},  244  (2006).

\bibitem{MetzlerKla2000PhR}
R. {Metzler} and J. {Klafter}, \physrep {\bf 339},  1  (2000).

\bibitem{MuirBacoli2004}
R. {Wang}, P. {Keast}, and P. {Muir}, Journal of Computational and Applied
  Mathematics {\bf 169},  127  (2004).
  
\bibitem{MDS10PPCF}
M.~A. {Malkov}, P.~H. {Diamond}, and R.~Z. {Sagdeev}, Plasma Physics and
  Controlled Fusion {\bf 52},  124006  (2010).

\bibitem{MetalEsc13}
M.~A. {Malkov} {\it et~al.}, \apj {\bf 768},  73  (2013).

\bibitem{NavaGabici2016}
L. {Nava} {\it et~al.}, \mnras {\bf 461},  3552  (2016).

\bibitem{DAngeloBlasi2016}
M. {D'Angelo}, P. {Blasi}, and E. {Amato}, \prd {\bf 94},  083003  (2016).

\bibitem{Gombosi93}
T.~I. {Gombosi} {\it et~al.}, \apj {\bf 403},  377  (1993).

\bibitem{Zank2000}
G.~P. {Zank}, J.~Y. {Lu}, W.~K.~M. {Rice}, and G.~M. {Webb}, Journal of Plasma
  Physics {\bf 64},  507  (2000).

\bibitem{Webb2000}
G.~M. {Webb}, M. {Pantazopoulou}, and G.~P. {Zank}, Journal of Physics A
  Mathematical General {\bf 33},  3137  (2000).

\bibitem{WebbTelegr06}
G.~M. Webb, G.~P. Zank, E.~K. Kaghashvili, and J.~A. le~Roux, The Astrophysical
  Journal {\bf 651},  211  (2006).

\bibitem{Fedorov2016}
Y.~I. {Fedorov} {\it et~al.}, Kinematics and Physics of Celestial Bodies {\bf
  32},  105  (2016).

\bibitem{Shalchi2006}
A. {Shalchi}, \aap {\bf 448},  809  (2006).  

\end{thebibliography}

\onecolumngrid

\selectlanguage{american}%

\selectlanguage{english}%
\twocolumngrid

\end{document}